\documentclass[onecolumn,amstext,amsmath,amssymb,prd,superscriptaddress,floatfix,showpacs,nofootinbib]{revtex4}
\usepackage[dvips]{epsfig}
\usepackage{slashed}
\usepackage{mathrsfs}
\usepackage{graphicx}
\usepackage{xcolor}
\usepackage{hyperref}
\usepackage{dcolumn}
\usepackage{bbold}
\usepackage{subfigure}
\usepackage{xspace}
\usepackage{bbm}
\usepackage{mathtools}
\usepackage{tabularx}

\newcolumntype{D}[1]{m{#1}}
\newcolumntype{C}[1]{>{\centering\arraybackslash}m{#1}}
\definecolor{heidelbeer}{rgb}{0.5,0,0.5}

\newcommand{\tr}[1]{\operatorname{Tr}\left[#1\right]}

\newcommand{\bracket}[3]{\big\langle#1\big|#2\big|#3\big\rangle}
\newcommand{\bra}[1]{\big\langle#1\big|}
\newcommand{\ket}[1]{\big|#1\big\rangle}

\graphicspath{
{./}
{./Figures/}
}
\begin{document}
\title{Real-time dynamics of open quantum spin systems driven by dissipative processes}

\author{F. Hebenstreit}
 \email[]{hebenstreit@itp.unibe.ch}
 \affiliation{Albert Einstein Center, Institute for Theoretical Physics, Bern University, 3012 Bern, Switzerland}

\author{D. Banerjee}
 \affiliation{NIC, DESY Zeuthen, Platanenallee 6, 15738 Zeuthen, Germany}

\author{M. Hornung}
 \affiliation{Albert Einstein Center, Institute for Theoretical Physics, Bern University, 3012 Bern, Switzerland}

\author{F.-J. Jiang}
 \affiliation{Department of Physics, National Taiwan Normal University, 88, Sec.~4, Ting-Chou Rd., Taipei 116, Taiwan}

\author{F. Schranz}
 \affiliation{Albert Einstein Center, Institute for Theoretical Physics, Bern University, 3012 Bern, Switzerland}

\author{U.-J. Wiese}
 \affiliation{Albert Einstein Center, Institute for Theoretical Physics, Bern University, 3012 Bern, Switzerland}


\begin{abstract}
We study the real-time evolution of large open quantum spin systems in two spatial dimensions, whose dynamics is entirely driven by a dissipative coupling to the environment.
We consider different dissipative processes and investigate the real-time evolution from an ordered phase of the Heisenberg or XY-model towards a disordered phase at late times, disregarding unitary Hamiltonian dynamics.
The corresponding Kossakowski-Lindblad equation is solved via an efficient cluster algorithm. 
We find that the symmetry of the dissipative process determines the time scales which govern the approach towards a new equilibrium phase at late times.
Most notably, we find a slow equilibration if the dissipative process conserves any of the magnetization Fourier modes.
In these cases, the dynamics can be interpreted as a diffusion process of the conserved quantity.
\end{abstract}
\pacs{03.65.Yz, 
      05.70.Ln, 
      75.10.Jm} 
\maketitle


\section{Introduction}

Simulating the real-time evolution of large quantum systems is one of the major challenges of modern theoretical physics, ranging from condensed matter physics to high-energy physics \cite{Stefanucci:2013,Berges:2004yj,Aoki:2014}.
Unlike quantum systems in thermal equilibrium for which quantum Monte Carlo simulations have proven to be an extremely powerful tool, there is no generally applicable approach to far-from-equilibrium quantum systems:
Exact diagonalization techniques become technically impossible for large systems due to the exponential growth of the Hilbert space with the system size.
On the other hand, the application of the quantum Monte Carlo method based on importance sampling fails for real-time simulations due to a severe sign or complex weight problem.
For gapped 1-dimensional systems with small entanglement, however, tensor network states underlying the density matrix renormalization group \cite{White:1992zz,Schollwock:2005zz} provide a good basis for simulating the real-time dynamics for moderate time intervals \cite{Cazalilla:2002,Vidal:2004,White:2004,Verstraete:2004,Daley:2004,Buyens:2013yza}. 
Moreover, simulations in the classical statistical or truncated Wigner approximation are valid in the limit of large occupation numbers which appear, for instance, at the early stages of relativistic heavy-ion collision or in the study of non-perturbative fermion production \cite{Polkovnikov:2010,Romatschke:2005pm,Gelis:2007kn,Gelis:2013rba,Berges:2013eia,Kasper:2014uaa}.
Finally, there have been attempts to apply stochastic quantization to real-time problems based on the complex Langevin equation even though the problem of run-away trajectories is not completely settled yet \cite{Parisi:1984cs,Huffel:1984mq,Berges:2006xc,Fukushima:2014iqa}.

One reason for the complexity of simulating the real-time evolution of large quantum systems on classical computers is the fact that isolated quantum systems tend to evolve into complicated entangled states such as those of Schr\"odinger's cat.
For this reason, it was proposed to use specifically designed quantum devices -- so-called quantum simulators -- to mimic quantum systems that are difficult to simulate classically due to the high degree of entanglement \cite{Feynman:1981tf}.
Due to the rapid development of atomic, molecular, and optical (AMO) physics in recent years, it has become possible to realize quantum simulators in systems of ultracold atoms \cite{Greiner:2002}.
As a consequence, applications in atomic and condensed matter physics \cite{Cirac:2008,Bloch:2008,Blatt:2008,AspuruGuzik:2008,Houck:2008,Lewenstein:2012} as well as in a particle physics context \cite{Kapit:2010qu,Zohar:2012ay,Banerjee:2012pg,Banerjee:2012xg,Zohar:2012xf,Tagliacozzo:2012vg,Tagliacozzo:2013,Zohar:2013zla,Wiese:2013uua,Kuhn:2014rha} are widely discussed.

Furthermore, real quantum systems are usually not isolated but coupled to a dissipative environment resulting in decoherence and the suppression of Schr\"odinger cat states.
Hence, it should be simpler to simulate quantum systems on a classical computer if we suppress the unitary Hamiltonian evolution and consider a pure, or at least sufficiently strong, dissipative coupling to an environment.
The corresponding dynamics is then governed by the Kossakowski-Lindblad equation \cite{Kossakowski:1972,Lindblad:1975ef}, which is the most general non-unitary, Markovian time-evolution equation for a density matrix which preserves the basic properties of Hermiticity and positive semi-definiteness.
The study of the interplay between the unitary Hamiltonian dynamics and the dissipative Markovian dynamics which may result in non-equilibrium (quantum) phase transitions has attracted much interest in recent years from both the theoretical and the experimental side \cite{Diehl:2008,Prozen:2008,Verstraete:2009,DallaTorre:2010,Diehl:2010,Muller:2012,Lesanovsky:2013,Sieberer:2013,Grandi:2013,Banchi:2014,Lang:2014}.

In this publication, we investigate the real-time evolution of large open quantum spin systems whose dynamics is entirely driven by specific randomized quantum measurements performed on pairs of neighboring spins.
The measurement processes give rise to a dissipative coupling to the environment, such that the system is described by the Kossakowski-Lindblad equation which can be solved with an efficient loop-cluster algorithm \cite{Evertz:1992rb,Wiese:1994}.
Unlike quantum trajectory methods, which are widely used in quantum optics to solve the Kossakowski-Lindblad equation and which scale exponentially with the size of the Hilbert space \cite{Daley:2014}, we can simulate the real-time evolution of the quantum system over arbitrarily long time intervals in any spatial dimension since the computational effort scales only linearly with the spatial volume.
We then find that the system is driven from an initial equilibrium state to a new equilibrium at final times.
Compared to a previous study of some of the authors \cite{Banerjee:2014yea}, we investigate a wider class of dissipative processes with different symmetries as well as using a variety of initial density matrices, corresponding to the $s=\frac{1}{2}$ Heisenberg model as well as the quantum XY model at low temperatures.
We emphasize that the associated Hamiltonians are only employed for preparing the initial density matrices but not for driving the subsequent real-time dynamics.

More specifically, we give a detailed account of the equilibration dynamics for the different dissipative processes.
We show that the symmetry of the specific process determines the time scales which govern the approach towards a new equilibrium phase at late times.
Most notably, we find a slow equilibration if the dissipative process conserves any of the magnetization Fourier modes.
In these cases, the dynamics can be interpreted as a diffusion process of the conserved quantity.

This paper is organized as follows:
In Secs.~\ref{sec:theory_sporadic} -- \ref{sec:theory_measurment} we recall the basic equations which govern the real-time evolution of a density matrix, most notably the Kossakowski-Lindblad equation, and introduce different measurement processes of pairs of neighboring spins.
In Sec.~\ref{sec:results_afm}, we discuss the real-time evolution of an anti-ferromagnetic Heisenberg model initial state subject to the different measurement processes in great detail and we analytically derive expectation values for the new equilibrium state at final times.
Conclusions and an outlook are presented in Sec.~\ref{sec:conclusion}.
We summarize the rules for forming loop-clusters for the different models and measurement processes in Appendix~\ref{app:cluster_rules}.
The real-time dynamics of a ferromagnetic Heisenberg model initial state and of a quantum XY-model initial state are discussed in Appendix~\ref{app:results_fm} and Appendix~\ref{app:results_xy}, respectively.


\section{Measurement-driven dissipative dynamics}

The real-time dynamics of a closed quantum system from an initial time $t_0$ to a final time $t_f$ is governed by the unitary time-evolution operator
\begin{equation}
 U(t_f,t_0)=\exp\left(-iH(t_f-t_0)\right)=U^\dagger(t_0,t_f) \ ,
\end{equation}
where we assumed a time-independent Hamiltonian operator $H$. 
Starting from an initial state which is specified by a (not necessarily thermal) density matrix $\rho(t_0)$, the expectation value of any observable $O$ at a later time is given by
\begin{equation}
 \label{eq:expectation}
 \big\langle O \big\rangle(t)=\tr{\rho(t)O}=\tr{\rho(t_0)U(t_0,t)OU(t,t_0)} \ ,
\end{equation}
where we employed the von-Neumann equation
\begin{equation}
 \label{eq:neumann}
 \rho(t)=U(t,t_0)\rho(t_0)U(t_0,t) \ .
\end{equation}
The expression \eqref{eq:expectation} may be rewritten as a real-time path integral along the Schwinger-Keldysh contour \cite{Schwinger:1960qe,Keldysh:1964ud} or, assuming a thermal density matrix $\rho(t_0)$, along the Konstantinov-Perel' contour \cite{Konstantinov:1961}.
The fact that the matrix elements of the time-evolution operator are in general complex leads to a severe sign problem, preventing the application of quantum Monte Carlo techniques.
Accordingly, we will consider a simpler situation in the following.

\subsection{Real-time evolution by sporadic measurements}
\label{sec:theory_sporadic}

We assume that certain observables $O_k$ are measured at intermediate times $t_k$ with $k\in\{1,\dotsc,N\}$, resulting in measurement results $o_k$.
The individual measurement results are associated with projection operators $P_{o_k}$, projecting on the subspace of the Hilbert space which is spanned by the eigenvectors of $O_k$ with eigenvalue $o_k$.
The projection operators are Hermitean $P_{o_k}=P^\dagger_{o_k}$, idempotent $P_{o_k}^2=P_{o_k}$, and fulfill $\sum_{o_k}P_{o_k}=\mathbb{1}$. 
We note that, given a certain density matrix $\rho(t_k)$ at time $t_k$, a measurement of the observable $O_k$ with result $o_k$ yields a new density matrix
\begin{equation}
 \label{eq:density}
 \rho'(t_k)=\frac{P_{o_k}\rho(t_k)P_{o_k}}{p(o_k)} \ ,
\end{equation}  
where $p(o_k)=\operatorname{tr}[\rho(t_k)P_{o_k}]$ denotes the probability of obtaining the measurement result $o_k$.
Alternating unitary real-time evolutions according to von-Neumann's equation \eqref{eq:neumann} and quantum measurements according to \eqref{eq:density} for a given initial density matrix $\rho(t_0)$ yields after $N$ steps
\begin{equation}
 \rho(t_N)=\frac{1}{p(o_1,o_2,\dotsc,o_N)}P_{o_N}U(t_N,t_{N-1})\dotsm P_{o_2}U(t_2,t_1)P_{o_1}U(t_1,t_0)\rho(t_0)U(t_0,t_1)P_{o_1}U(t_1,t_2)P_{o_2}\dotsm U(t_{N-1},t_N)P_{o_N} \ .
\end{equation}
Here, $p(o_1,o_2,\dotsc,o_N)$ denotes the probability of finding the measurement results $\{o_1,o_2,\dotsc,o_N\}$ upon starting with the initial density matrix
\begin{equation}
 \rho(t_0)=\sum_{i}p_i\ket{i}\bra{i} \ ,
\end{equation}
with $0\leq p_i\leq1$ and $\sum_{i}p_i=1$.
Accordingly, the probability of reaching a certain final state $\ket{f}$ at the final time $t_f$ after a sequence of $N$ measurements with measurement results $\{o_1,o_2,\dotsc,o_N\}$, is given by \cite{Griffiths:1984rx}
\begin{align}
 p_{\rho_0,f}&(o_1,o_2,\dotsc,o_N)=\bra{f}U(t_f,t_N)\rho(t_N)U(t_N,t_f)\ket{f}= \nonumber \\
 &\sum_{i}p_i\bra{i}U(t_0,t_1)P_{o_1}U(t_1,t_2)P_{o_2}\dotsm P_{o_N}U(t_N,t_{f})\ket{f}\bra{f}U(t_f,t_N)P_{o_N}\dotsm P_{o_2}U(t_2,t_1)P_{o_1}U(t_1,t_0)\ket{i} \ .
\end{align}
This transition probability is still plagued by a severe sign problem.
In order to make the problem amenable to Monte Carlo importance sampling, we further assume in this publication that the real-time dynamics of the quantum system is entirely driven by the measurement process, i.~e. $U(t_{k+1},t_k)=\mathbb{1}$.
The more challenging step of the real-time evolution being driven by both a non-trivial time-evolution operator and by quantum measurements is currently under investigation but beyond the scope of this publication. 
It has to be emphasized, however, that even this simplified problem results in interesting highly non-trivial quantum dynamics \cite{Banerjee:2014yea}, as discussed in more detail in the following sections. 

In this case, the probability of reaching a final state $\ket{f}$ at the final time $t_f\equiv t_N$ after a sequence of $N$ measurements with measurement results $\{o_1,\dotsc,o_N\}$ is calculated according to
\begin{align}
 p_{\rho_0,f}(o_1,\dotsc,o_N) = \sum_{i}p_i\bra{i}P_{o_1}\dotsm P_{o_N}\ket{f}\bra{f}P_{o_N}\dotsm P_{o_1}\ket{i} \ .
\end{align}
In order to derive a real-time path integral along the Schwinger-Keldysh contour leading from the initial time $t_0$ to the final time $t_N$ and back, we insert complete sets of states between the individual projection operators.
Denoting these complete sets by $\{\ket{n'_k}\}$ on the forward branch and by $\{\ket{n_k}\}$ on the backward branch, with $k\in\{1,\dotsc,N-1\}$, the resolution of the identity operator reads $\sum_{n_k}\ket{n_k}\bra{n_k}=\sum_{n_k^\prime}\ket{n_k^\prime}\bra{n_k^\prime}=\mathbb{1}$.
Accordingly, we find
\begin{equation}
 \label{eq:sk_path_nosum}
 p_{\rho_0,f}(o_1,\dotsc,o_N)=\sum_{i}p_i\sum_{n_1,n_1'}\cdots\sum_{n_{N-1},n'_{N-1}}\prod_{k=1}^{N}{\bracket{n_{k-1}n'_{k-1}}{P_{o_k}\otimes P^*_{o_k}}{n_{k}n'_{k}}} \ ,
\end{equation}
where we introduced the notation
\begin{equation}
 \bracket{n_{k-1}n'_{k-1}}{P_{o_k}\otimes P^*_{o_k}}{n_kn'_k}\equiv\bracket{n_{k-1}}{P_{o_k}}{n_k}\bracket{n'_{k-1}}{P_{o_k}}{n'_k}^* \ ,
\end{equation}
with $\bra{n_0n_0'}\equiv\bra{ii}$ and $\ket{n_Nn'_N}\equiv\ket{ff}$.
We note that the doubled Hilbert space of states $\{\ket{n_kn'_k}\}$ encompasses both branches of the Schwinger-Keldysh contour.

In the following, we will be mainly interested in the total probability of reaching the final state $\ket{f}$ irrespective of the intermediate measurement results $\{o_1,o_2,\dotsc,o_N\}$, which is calculated according to
\begin{equation}
 \label{eq:sk_path_sum}
 p_{\rho_0,f}=\sum_{o_1}\dotsm\sum_{o_N}p_{\rho_0,f}(o_1,\dotsc,o_N)=\sum_{i}p_i\sum_{n_1,n_1'}\cdots\sum_{n_{N-1},n'_{N-1}}\prod_{k=1}^{N}{\bracket{n_{k-1}n'_{k-1}}{\widetilde{P}_k}{n_kn'_k}} \ .
\end{equation}
Here, 
\begin{equation}
 \widetilde{P}_k=\sum_{o_k}P_{o_k}\otimes P_{o_k}^* 
\end{equation}
is obtained by summing over all possible measurement results $o_k$ at time $t_k$.
The final density matrix $\rho(t_N)$, which is the result of the real-time dynamics generated by the sequence of $N$ measurements, is then given by
\begin{equation}
 \label{eq:sporadic}
 \rho(t_N)=\sum_{o_1}\dotsc\sum_{o_N}P_{o_N}\dotsm P_{o_1}\rho(t_0)P_{o_1}\dotsm P_{o_N} \ , 
\end{equation}
with $\tr{\rho(t_N)}=\sum_{f}p_{\rho_0,f}=1$.

\subsection{Lindblad evolution}
\label{sec:theory_lindblad}

In the previous section we described the real-time evolution of the density matrix if the dynamics is entirely driven by a sporadic measurement process at discrete times $t_k$.
In the following, we consider the continuous time limit $t_{k+1}-t_k=\epsilon\to0$ and suppose that the measurement happens only with a certain probability per unit of time. 
Accordingly, the quantum system can then be considered as being continuously monitored by the environment.

This situation can be described by the Kossakowski-Lindblad equation \cite{Kossakowski:1972,Lindblad:1975ef}, which is the most general non-unitary, Markovian time evolution equation of a density matrix which preserves the basic properties of Hermiticity and positive semi-definiteness.
It is characterized by a set of Lindblad operators, describing the possible quantum jumps the system may undergo at any instant of time
\begin{equation}
 L_{o_k}=\sqrt{\epsilon\gamma}P_{o_k} \ ,
\end{equation}
which obey
\begin{equation}
 \left(1-\epsilon\gamma \right)\mathbb{1}+\sum_{o_{k}}{L^\dagger_{o_k}L_{o_k}}=\mathbb{1} \ .
\end{equation}
The parameter $\gamma$ determines the probability per unit time that a certain quantum jump occurs.
Using the notation of the previous section, a certain quantum jump operator $L_{o_k}$ corresponds to the measurement of an observable $O_k$ yielding the measurement result $o_k$.\footnote{In general, the quantum system may interact in different ways with the environment so that we have to include several different observables $O_k$ at any instant of time. 
For simplicity, we restrict ourselves to a single observable in this section and discuss the more general situation in Sec.~\ref{sec:theory_large_systems}.}
The Kossakowski-Lindblad equation is then given by
\begin{equation}
 \frac{d}{dt}\rho(t)=\frac{1}{\epsilon}\sum_{o_{k}}\left[L_{o_k}\rho(t)L^\dagger_{o_k}-\frac{1}{2}L^\dagger_{o_k}L_{o_k}\rho(t)-\frac{1}{2}\rho(t)L^\dagger_{o_k}L_{o_k}\right]=\gamma\sum_{o_k}\left[P_{o_k}\rho(t)P_{o_k}-\rho(t)\right] \ .
\end{equation}
We emphasize the absence of a unitary time evolution term $-i[H,\rho(t)]$ on the right-hand side as we assume that the real-time dynamics is entirely driven by the dissipative process.

We translate this continuous-time equation back to discrete times since we employ a discrete-time algorithm in our numerical simulations.
We note, however, that the systems could also be simulated directly in continuous time \cite{Beard:1996wj}.  
To this end, we recursively perform a forward finite-difference discretization of the time derivative to obtain
\begin{equation}
 \label{eq:recursion}
 \rho(t_k)=(1-\epsilon\gamma)\rho(t_{k-1}) +\epsilon\gamma\sum_{o_k}P_{o_k}\rho(t_{k-1})P_{o_k} \ ,
\end{equation}
where we disregard higher discretization errors. 
Here, $k\in\{1,\dotsc,N\}$ labels the discrete times $t_k$ at which the quantum system potentially interacts with the environment.
The recursive equation \eqref{eq:recursion} allows us to express the final density matrix $\rho(t_N)$ as a function of the initial density matrix $\rho(t_0)$ according to
\begin{equation}
 \rho(t_N)=(1-\epsilon\gamma)^N\rho(t_0)+\sum_{s=1}^{N}(\epsilon\gamma)^s(1-\epsilon\gamma)^{N-s}\sum_{i_1<\dotsc<i_{s}}\sum_{o_{i_1}}\dotsm\sum_{o_{i_{s}}}P_{o_{i_{s}}}\dotsm P_{o_{i_1}}\rho(t_0)P_{o_{i_1}}\dotsm P_{o_{i_{s}}} \ .
\end{equation}
The individual terms have simple interpretations: 
The first term $\sim(1-\epsilon\gamma)^N$ correspond to a time history without any interactions whereas the remaining terms all include measurements.
In fact, the order $s$ determines how often the quantum system interacts with the environment.
Most notably, the last term $\sim(\epsilon\gamma)^N$ corresponds to a time history with quantum jumps at every discrete time step.

Following the same construction as in the previous section, we can again derive a real-time path integral along the Schwinger-Keldysh contour leading from the initial time $t_0$ to the final time $t_N$ and back.
Most notably, the total probability of reaching a final state $\ket{f}$ irrespective of any possible intermediate measurement results is given by
\begin{equation}
 \label{eq:Lindblad}
 p_{\rho_0,f}=\sum_{i}p_i\sum_{n_1,n_1'}\cdots\sum_{n_{N-1},n'_{N-1}}\prod_{k=1}^{N}{\bracket{n_{k-1}n'_{k-1}}{(1-\epsilon\gamma)\mathbb{1}\otimes\mathbb{1}+\epsilon\gamma\widetilde{P}_k}{n_kn'_k}} \ .
\end{equation}
As a matter of fact, we recover the result for the real-time evolution due to sporadic measurements \eqref{eq:sk_path_sum} in the limit $\epsilon\gamma=1$, corresponding to a measurement at every time step with probability $1$.

\subsection{Measurement processes}
\label{sec:theory_measurment}

In this study, we investigate 2-dimensional systems of quantum spins $s=\tfrac{1}{2}$.
In order to generate dissipative dynamics, we apply a given measurement process corresponding to a specific observable $O_k$.
As we will see, different measurement processes result in different dynamic behavior.

In order to set the stage, we start with the simple situation of just two quantum spins at positions $x$ and $y$.
At the end of this section, we will generalize the two-spin system to a large 2-dimensional system, appreciating that this generalization is feasible for any dimension. 
In the following, we take the $3$-direction as the quantization axis and denote the two eigenvalues of $S_x^3$ by $s_x=\,\uparrow\,=\tfrac{1}{2}$ and $s_x=\,\downarrow\,=-\tfrac{1}{2}$.
The Hilbert space of the two-spin system is composed of four states and denoted by
\begin{equation}
 \label{eq:hilbert}
 \mathcal{H}=\left\{\ket{s_x\,s_y}\right\}=\left\{\ket{\uparrow\,\uparrow},\ket{\uparrow\,\downarrow},\ket{\downarrow\,\uparrow},\ket{\downarrow\,\downarrow}\right\} \ .
\end{equation}

\subsubsection{Measurement process: $\vec{S}^2$}

The first measurement process under consideration corresponds to the total spin of the system
\begin{equation}
 O^{(1)}=\vec{S}^2=(\vec{S}_x+\vec{S}_y)^2=\vec{S}_x^2+\vec{S}_y^2+2\vec{S}_x\cdot\vec{S}_y=\frac{3}{2}\mathbb{1}+2\vec{S}_x\cdot\vec{S}_y \ .
\end{equation}
This operator conserves each of the components of the spin vector $\vec{S}=\vec{S}_x+\vec{S}_y$ so that
\begin{equation}
 \label{eq:cons_tot}
 [\vec{S}^2,S^i]=0 \ .
\end{equation}
The simultaneous eigenstates of $\vec{S}^2$ and, for instance, its $3$-component $S^3=S_x^3+S_y^3$ are denoted by $\ket{SS^3}$.
They are given by the singlet state
\begin{equation}
 \ket{00}=\tfrac{1}{\sqrt{2}}\left(\ket{\uparrow\,\downarrow}-\ket{\downarrow\,\uparrow}\right) \ ,
\end{equation}
and the triplet states
\begin{equation}
 \left\{\ket{11}=\ket{\uparrow\,\uparrow} \ ,\ \ket{10}=\tfrac{1}{\sqrt{2}}\left(\ket{\uparrow\,\downarrow}+\ket{\downarrow\,\uparrow}\right) \ , \ \ket{1-1}=\ket{\downarrow\,\downarrow}\right\} \ .
\end{equation}
The projection operators on the different measurement results $S=0$ and $S=1$, respectively, are then given by
\begin{subequations}
\begin{align}
 P_0&=\ket{00}\bra{00} \ , \\
 P_1&=\ket{11}\bra{11}+\ket{10}\bra{10}+\ket{1-1}\bra{1-1} \ .
\end{align}
\end{subequations}
In the basis \eqref{eq:hilbert}, these projection operators have the matrix representation
\begin{equation}
 \label{eq:proj_total}
 \hat{P}_0=\frac{1}{2}\begin{pmatrix}0&0&0&0\\0&1&-1&0\\0&-1&1&0\\0&0&0&0\end{pmatrix} \quad , \quad \hat{P}_1=\frac{1}{2}\begin{pmatrix}2&0&0&0\\0&1&1&0\\0&1&1&0\\0&0&0&2\end{pmatrix} \ .
\end{equation}
We note that certain matrix elements for measuring $S=0$ in the doubled Hilbert space of states $\ket{n_kn'_k}\equiv\ket{s_{x,k}s_{y,k}s'_{x,k}s'_{y,k}}$ are negative 
\begin{align}
 \label{eq:sign_total}
 &\bracket{s_{x,k}s_{y,k}s'_{x,k}s'_{y,k}}{P_0\otimes P^*_0}{s_{x,k+1}s_{y,k+1}s'_{x,k+1}s'_{y,k+1}}=\nonumber \\
 &\qquad\frac{1}{4}\left(\ \delta_{s_{x,k},s_{x,k+1}}\delta_{s_{y,k},s_{y,k+1}}\delta_{s'_{x,k},s'_{x,k+1}}\delta_{s'_{y,k},s'_{y,k+1}}+
 \delta_{s_{x,k},s_{y,k+1}}\delta_{s_{y,k},s_{x,k+1}}\delta_{s'_{x,k},s'_{y,k+1}}\delta_{s'_{y,k},s'_{x,k+1}}\right.\nonumber \\
 &\qquad\quad\left.-\delta_{s_{x,k},s_{x,k+1}}\delta_{s_{y,k},s_{y,k+1}}\delta_{s'_{x,k},s'_{y,k+1}}\delta_{s'_{y,k},s'_{x,k+1}}-
 \delta_{s_{x,k},s_{y,k+1}}\delta_{s_{y,k},s_{x,k+1}}\delta_{s'_{x,k},s'_{x,k+1}}\delta_{s'_{y,k},s'_{y,k+1}}\right) \ ,
\end{align}
giving rise to a sign problem in the corresponding real-time path integral \eqref{eq:sk_path_nosum}.
On the other hand, the matrix elements for measuring $S=1$ are always non-negative
\begin{align}
 &\bracket{s_{x,k}s_{y,k}s'_{x,k}s'_{y,k}}{P_1\otimes P^*_1}{s_{x,k+1}s_{y,k+1}s'_{x,k+1}s'_{y,k+1}}=\nonumber \\
 &\qquad\frac{1}{4}\left(\ \delta_{s_{x,k},s_{x,k+1}}\delta_{s_{y,k},s_{y,k+1}}\delta_{s'_{x,k},s'_{x,k+1}}\delta_{s'_{y,k},s'_{y,k+1}}+
 \delta_{s_{x,k},s_{y,k+1}}\delta_{s_{y,k},s_{x,k+1}}\delta_{s'_{x,k},s'_{y,k+1}}\delta_{s'_{y,k},s'_{x,k+1}}\right.\nonumber \\
 &\qquad\quad\left.+\delta_{s_{x,k},s_{x,k+1}}\delta_{s_{y,k},s_{y,k+1}}\delta_{s'_{x,k},s'_{y,k+1}}\delta_{s'_{y,k},s'_{x,k+1}}+
 \delta_{s_{x,k},s_{y,k+1}}\delta_{s_{y,k},s_{x,k+1}}\delta_{s'_{x,k},s'_{x,k+1}}\delta_{s'_{y,k},s'_{y,k+1}}\right) \ .
\end{align}
Remarkably, the sign problem arising from \eqref{eq:sign_total} is eliminated by averaging over both measurement results 
\begin{align}
 \label{eq:cluster_total}
 &\bracket{s_{x,k}s_{y,k}s'_{x,k}s'_{y,k}}{P_0\otimes P^*_0+P_1\otimes P^*_1}{s_{x,k+1}s_{y,k+1}s'_{x,k+1}s'_{y,k+1}}=\nonumber \\
 &\qquad\frac{1}{2}\left(\ \delta_{s_{x,k},s_{x,k+1}}\delta_{s_{y,k},s_{y,k+1}}\delta_{s'_{x,k},s'_{x,k+1}}\delta_{s'_{y,k},s'_{y,k+1}}+
 \delta_{s_{x,k},s_{y,k+1}}\delta_{s_{y,k},s_{x,k+1}}\delta_{s'_{x,k},s'_{y,k+1}}\delta_{s'_{y,k},s'_{x,k+1}}\right)\geq0 \ ,
\end{align}
so that the corresponding real-time path integral \eqref{eq:sk_path_sum} is not plagued by a sign problem anymore.

\subsubsection{Measurement process: $S_x^1S_y^1$}

The second measurement process under consideration corresponds to a measurement of the products of $1$-components of the quantum spins
\begin{equation}
 O^{(2)}=S_x^1S_y^1 \ .
\end{equation}
Accordingly, this measurement distinguishes whether the $1$-components of the two spins are the same or different.
Equivalently, we could have also chosen a measurement process $S_x^2S_y^2$ regarding the $2$-components of the two spins.
As we choose the $3$-direction as the quantization axis \eqref{eq:hilbert}, the two eigenstates of $O^{(2)}$ corresponding to parallel spins are denoted by
\begin{equation}
 \left\{\ket{\parallel_1}=\tfrac{1}{\sqrt{2}}\left(\ket{\uparrow\,\uparrow}+\ket{\downarrow\,\downarrow}\right) \ ,\ \ket{\parallel_2}=\tfrac{1}{\sqrt{2}}\left(\ket{\uparrow\,\downarrow}+\ket{\downarrow\,\uparrow}\right) \right\} \ ,
\end{equation}
whereas the two eigenstates corresponding to anti-parallel spins are given by
\begin{equation}
 \left\{\ket{\nparallel_1}=\tfrac{1}{\sqrt{2}}\left(\ket{\uparrow\,\uparrow}-\ket{\downarrow\,\downarrow}\right) \ ,\ \ket{\nparallel_2}=\tfrac{1}{\sqrt{2}}\left(\ket{\uparrow\,\downarrow}-\ket{\downarrow\,\uparrow}\right) \right\} \ .
\end{equation}
The projection operators on parallel and anti-parallel $1$-components of the two spins, respectively, are then calculated according to
\begin{subequations}
\begin{align}
 P_\parallel&=\ket{\parallel_1}\bra{\parallel_1}+\ket{\parallel_2}\bra{\parallel_2} \ , \\
 P_\nparallel&=\ket{\nparallel_1}\bra{\nparallel_1}+\ket{\nparallel_2}\bra{\nparallel_2} \ .
\end{align}
\end{subequations}
Again, expressing these projection operators in the basis \eqref{eq:hilbert}, we obtain
\begin{equation}
 \hat{P}_\parallel=\frac{1}{2}\begin{pmatrix}1&0&0&1\\0&1&1&0\\0&1&1&0\\1&0&0&1\end{pmatrix} \quad , \quad  \hat{P}_\nparallel=\frac{1}{2}\begin{pmatrix}1&0&0&-1\\0&1&-1&0\\0&-1&1&0\\-1&0&0&1\end{pmatrix} \ .
\end{equation}
Like before, certain components of $\hat{P}_\nparallel$ are negative and give rise to a sign problem in the real-time path integral \eqref{eq:sk_path_nosum}.
On the other hand, the matrix elements for measuring parallel $1$-components are still non-negative.
Specifically, we have
\begin{subequations}
\begin{align}
 \label{eq:sign_x}
 &\bracket{s_{x,k}s_{y,k}s'_{x,k}s'_{y,k}}{P_\nparallel\otimes P^*_\nparallel}{s_{x,k+1}s_{y,k+1}s'_{x,k+1}s'_{y,k+1}}=\nonumber \\
 &\qquad\frac{1}{4}\left(\ \delta_{s_{x,k},s_{x,k+1}}\delta_{s_{y,k},s_{y,k+1}}\delta_{s'_{x,k},s'_{x,k+1}}\delta_{s'_{y,k},s'_{y,k+1}}+
 \delta_{s_{x,k},-s_{x,k+1}}\delta_{s_{y,k},-s_{y,k+1}}\delta_{s'_{x,k},-s'_{x,k+1}}\delta_{s'_{y,k},-s'_{y,k+1}}\right.\nonumber \\
 &\qquad\quad\left.-\delta_{s_{x,k},s_{x,k+1}}\delta_{s_{y,k},s_{y,k+1}}\delta_{s'_{x,k},-s'_{x,k+1}}\delta_{s'_{y,k},-s'_{y,k+1}}-
 \delta_{s_{x,k},-s_{x,k+1}}\delta_{s_{y,k},-s_{y,k+1}}\delta_{s'_{x,k},s'_{x,k+1}}\delta_{s'_{y,k},s'_{y,k+1}}\right) \ , \\
 &\bracket{s_{x,k}s_{y,k}s'_{x,k}s'_{y,k}}{P_\parallel\otimes P^*_\parallel}{s_{x,k+1}s_{y,k+1}s'_{x,k+1}s'_{y,k+1}}=\nonumber \\
 &\qquad\frac{1}{4}\left(\ \delta_{s_{x,k},s_{x,k+1}}\delta_{s_{y,k},s_{y,k+1}}\delta_{s'_{x,k},s'_{x,k+1}}\delta_{s'_{y,k},s'_{y,k+1}}+
 \delta_{s_{x,k},-s_{x,k+1}}\delta_{s_{y,k},-s_{y,k+1}}\delta_{s'_{x,k},-s'_{x,k+1}}\delta_{s'_{y,k},-s'_{y,k+1}}\right.\nonumber \\
 &\qquad\quad\left.+\delta_{s_{x,k},s_{x,k+1}}\delta_{s_{y,k},s_{y,k+1}}\delta_{s'_{x,k},-s'_{x,k+1}}\delta_{s'_{y,k},-s'_{y,k+1}}+
 \delta_{s_{x,k},-s_{x,k+1}}\delta_{s_{y,k},-s_{y,k+1}}\delta_{s'_{x,k},s'_{x,k+1}}\delta_{s'_{y,k},s'_{y,k+1}}\right) \ .
\end{align}
\end{subequations}
As before, the sign problem arising from \eqref{eq:sign_x} is completely eliminated by averaging over both measurement results
\begin{align}
 \label{eq:cluster_x}
 &\bracket{s_{x,k}s_{y,k}s'_{x,k}s'_{y,k}}{P_\parallel\otimes P^*_\parallel+P_\nparallel\otimes P^*_\nparallel}{s_{x,k+1}s_{y,k+1}s'_{x,k+1}s'_{y,k+1}}=\nonumber \\
 &\qquad\quad\frac{1}{2}\left(\ \delta_{s_{x,k},s_{x,k+1}}\delta_{s_{y,k},s_{y,k+1}}\delta_{s'_{x,k},s'_{x,k+1}}\delta_{s'_{y,k},s'_{y,k+1}}+
 \delta_{s_{x,k},-s_{x,k+1}}\delta_{s_{y,k},-s_{y,k+1}}\delta_{s'_{x,k},-s'_{x,k+1}}\delta_{s'_{y,k},-s'_{y,k+1}}\right)\geq0 \ .
\end{align}

\subsubsection{Measurement process: $S_x^+S_y^+ + S_x^-S_y^-$}

Finally, the last measurement process under consideration corresponds to the observable
\begin{equation}
 O^{(3)}=S_x^+S_y^+ + S_x^-S_y^-=2(S_x^1S_y^1-S_x^2S_y^2) \ ,
\end{equation}
where we introduced the raising and lowering operators according to $S_x^{\pm}=S_x^1\pm iS_x^{2}$.
The intriguing feature of this observable is that it conserves the difference of the $3$-components of the two spins
\begin{equation}
 \label{eq:cons_rl}
 [S_x^+S_y^+ + S_x^-S_y^-,S_x^3-S_y^3]=0 \ .
\end{equation}
This process measures the correlation between parallelism and anti-parallelism of the $1$-components and $2$-components of the two quantum spins.
Given that the $1$-components are the same whereas the $2$-components are different, the eigenvalue of $O^{(3)}$ is $+1$ and the corresponding eigenvector is given by 
\begin{equation}
 \ket{+}=\tfrac{1}{\sqrt{2}}\left(\ket{\uparrow\,\uparrow}+\ket{\downarrow\,\downarrow}\right) \ .
\end{equation}
In the opposite case, where the $1$-components are different but the $2$-components are the same, the eigenvalue of $O^{(3)}$ is $-1$ and its eigenvector reads
\begin{equation}
 \ket{-}=\tfrac{1}{\sqrt{2}}\left(\ket{\uparrow\,\uparrow}-\ket{\downarrow\,\downarrow}\right) \ .
\end{equation}
Finally, there are two degenerate eigenvalues $0$ corresponding to the case where the $1$-components and $2$-components are both either the same or different. 
Because of \eqref{eq:cons_rl}, we can further distinguish these states by their simultaneous eigenvalue of $S_x^3-S_y^3$ and we denote them by
\begin{equation}
 \left\{ \ket{0_{+}}=\ket{\uparrow\,\downarrow} \ ,\ \ket{0_{-}}=\ket{\downarrow\,\uparrow} \right\} \ .
\end{equation}
The three projection operators on the different measurement results are thus given by
\begin{subequations}
\begin{align}
 P_{+}&=\ket{+}\bra{+} \ , \\
 P_0&=\ket{0_+}\bra{0_+} + \ket{0_-}\bra{0_-} \ , \\
 P_{-}&=\ket{-}\bra{-} \ .
\end{align}
\end{subequations}
Equivalently, the matrix representation in the basis \eqref{eq:hilbert} reads
\begin{equation}
 \hat{P}_{+}=\frac{1}{2}\begin{pmatrix}1&0&0&1\\0&0&0&0\\0&0&0&0\\1&0&0&1\end{pmatrix} \quad , \quad  \hat{P}_0=\begin{pmatrix}0&0&0&0\\0&1&0&0\\0&0&1&0\\0&0&0&0\end{pmatrix} \quad , \quad \hat{P}_{-}=\frac{1}{2}\begin{pmatrix}1&0&0&-1\\0&0&0&0\\0&0&0&0\\-1&0&0&1\end{pmatrix} \ .
\end{equation}
Like before, the negative entries in $\hat{P}_{-}$ give rise to a sign problem in the real-time path integral \eqref{eq:sk_path_nosum}, whereas all remaining matrix elements are non-negative
\begin{subequations}
\begin{align}
 \label{eq:sign_rl}
 &\bracket{s_{x,k}s_{y,k}s'_{x,k}s'_{y,k}}{P_-\otimes P^*_-}{s_{x,k+1}s_{y,k+1}s'_{x,k+1}s'_{y,k+1}}=\nonumber \\
 &\qquad \qquad \qquad 4s_{x,k}s_{x,k+1}s'_{x,k}s'_{x,k+1}\delta_{s_{x,k},s_{y,k}}\delta_{s_{x,k+1},s_{y,k+1}}\delta_{s'_{x,k},s'_{y,k}}\delta_{s'_{x,k+1},s'_{y,k+1}} \ , \\
 &\bracket{s_{x,k}s_{y,k}s'_{x,k}s'_{y,k}}{P_+\otimes P^*_+}{s_{x,k+1}s_{y,k+1}s'_{x,k+1}s'_{y,k+1}}=\nonumber \\
 &\qquad \qquad \qquad \frac{1}{4}\delta_{s_{x,k},s_{y,k}}\delta_{s_{x,k+1},s_{y,k+1}}\delta_{s'_{x,k},s'_{y,k}}\delta_{s'_{x,k+1},s'_{y,k+1}}\ \ , \\
 &\bracket{s_{x,k}s_{y,k}s'_{x,k}s'_{y,k}}{P_0\otimes P^*_0}{s_{x,k+1}s_{y,k+1}s'_{x,k+1}s'_{y,k+1}}=\nonumber \\
 &\qquad \qquad \qquad \frac{(1+4s_{x,k}s_{x,k+1})(1+4s'_{x,k}s'_{x,k+1})}{4}\delta_{s_{x,k},-s_{y,k}}\delta_{s_{x,k+1},-s_{y,k+1}}\delta_{s'_{x,k},-s'_{y,k}}\delta_{s'_{x,k+1},-s'_{y,k+1}} \ .
\end{align}
\end{subequations}
The sign problem arising from \eqref{eq:sign_rl} is again completely eliminated by averaging over all three measurement results
\begin{align}
 \label{eq:cluster_rl}
 &\bracket{s_{x,k}s_{y,k}s'_{x,k}s'_{y,k}}{P_+\otimes P^*_++P_0\otimes P^*_0+P_-\otimes P^*_-}{s_{x,k+1}s_{y,k+1}s'_{x,k+1}s'_{y,k+1}}= \nonumber \\
 &\qquad \qquad \qquad4s_{x,k}s'_{x,k}\left(s_{x,k}s'_{x,k}+s_{x,k+1}s'_{x,k+1}\right)\delta_{s_{x,k},s_{y,k}}\delta_{s_{x,k+1},s_{y,k+1}}\delta_{s'_{x,k},s'_{y,k}}\delta_{s'_{x,k+1},s'_{y,k+1}}\ + \nonumber \\
 &\qquad \qquad \qquad4s_{x,k}s'_{x,k}\big(s_{x,k}+s_{x,k+1}\big)\left(s'_{x,k}+s'_{x,k+1}\right)\delta_{s_{x,k},-s_{y,k}}\delta_{s_{x,k+1},-s_{y,k+1}}\delta_{s'_{x,k},-s'_{y,k}}\delta_{s'_{x,k+1},-s'_{y,k+1}}\geq0 \ . 
\end{align}

\subsubsection{Generalization to large higher dimensional systems}
\label{sec:theory_large_systems}

We now generalize the results for the two-spin system to large $2$-dimensional systems of quantum spins $s=\tfrac{1}{2}$ on bipartite square lattices of the size $L\times L$ with periodic boundary conditions.
In fact, the generalization to any number of dimensions is feasible and straightforward.
To this end, we assume that any measurement process affects only nearest-neighbor quantum spin pairs.
Accordingly, we can take advantage of the results of the previous section for the two-spin system.

\begin{figure}[t]
 \includegraphics[width=0.5\columnwidth]{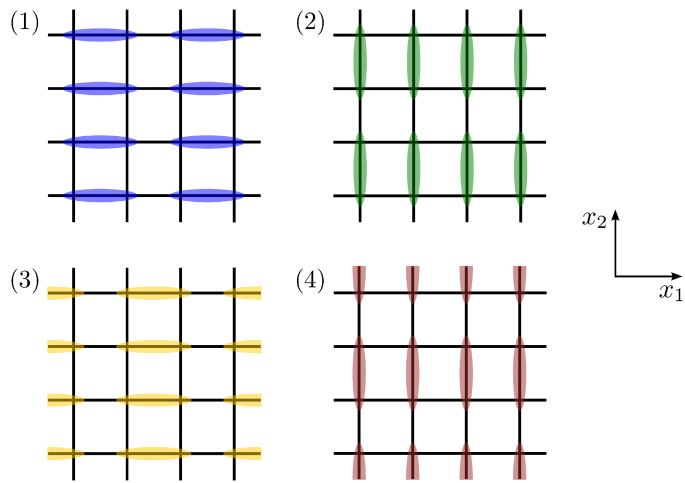}
 \caption{\label{fig:discretization}
 [Color online] Visualization of the discretization of the physical system.
 The oval-shaped objects denote the nearest-neighbor spin pairs which may interact at discretization step $1$ -- $4$, respectively.}
\end{figure}

Specifically, in order to allow for an efficient implementation of the real-time evolution, we discretize the system in the following way (cf.~Fig.~\ref{fig:discretization}): 
In a first step, all pairs of neighboring spins separated in the $1$-direction at $x=(x_1,x_2)$ and $y=(x_1+1,x_2)$ with even $x_1$ can be measured.
The second step allows for measuring spin pairs in the $2$-direction at $x=(x_1,x_2)$ and $y=(x_1,x_2+1)$ with even $x_2$.
In the third and fourth step, the measurement process affects spins with odd $x_1$ and odd $x_2$, respectively.
Accordingly, all possible interactions between nearest-neighbor spins occur during these four steps, which can then be repeated an arbitrary number of times.

For a sporadic measurement process, all of the neighboring spins are measured simultaneously according to the four-step scheme. 
The corresponding probability of reaching the final state $\ket{f}$ irrespective of the intermediate measurement results \eqref{eq:sk_path_sum} is then given by
\begin{equation}
 p_{\rho_0,f}=\sum_{i}p_i\sum_{n_1,n_1'}\cdots\sum_{n_{N-1},n'_{N-1}}\prod_{k=1}^{N}\prod_{\langle xy\rangle}{\bracket{s_{x,k-1}s_{y,k-1}s'_{x,k-1}s'_{y,k-1}}{\widetilde{P}_{k,xy}}{s_{x,k}s_{y,k}s'_{x,k}s'_{y,k}}} \ ,
\end{equation}
where $n_k\equiv[s_{x,k}]$ represents the $2^{N^2}$ possible spin states.
We emphasize that the $2N^2$ projectors $\widetilde{P}_{k,xy}$ are chosen according to the aforementioned four-step scheme, i.~e.~for $k\in\{1,5,9,...\}$ only those $N^2/2$ projection operators are used which connect neighboring spins separated in the $1$-direction with even $x_1$, and likewise for the other three steps.

In the following, we study the more natural situation where the measurement process does not affect all quantum spins simultaneously but only a small number of randomly chosen neighboring spins.
This situation is again described by the Kossakowski-Lindblad equation, which now results in
\begin{equation}
 \label{eq:Lindblad_pairs}
 p_{\rho_0,f}=\sum_{i}p_i\sum_{n_1,n_1'}\cdots\sum_{n_{N-1},n'_{N-1}}\prod_{k=1}^{N}\prod_{\langle xy\rangle}{\bracket{s_{x,k-1}s_{y,k-1}s'_{x,k-1}s'_{y,k-1}}{(1-\epsilon\gamma)\mathbb{1}\otimes\mathbb{1}+\epsilon\gamma\widetilde{P}_{k,xy}}{s_{x,k}s_{y,k}s'_{x,k}s'_{y,k}}} \ .
\end{equation}
In fact, we retain the aforementioned four-step scheme for discretizing the system for our numerical simulations.
We note, however, that four discrete time steps are considered as one physical time step in the spirit of a Suzuki-Trotter decomposition such that $t_k\gamma=\epsilon\gamma k/4$.
Moreover, we emphasize that the particular order of the four-step scheme is irrelevant for the Lindblad evolution since the interaction is anyway randomized in the continuous time limit.
By taking the limit $\epsilon\gamma\to0$, only a small number of randomly chosen neighboring spin pairs interact at any instant of time.

The physical process under consideration is the following:
Starting from a thermal initial density matrix $\rho(t_0)$, the measurement process induces a non-trivial real-time evolution, resulting in a final density matrix $\rho(t_N)$.
In order to define $\rho(t_0)$, we will choose different model Hamiltonians $H$ corresponding to the $s=\tfrac{1}{2}$ anti-ferromagnetic (AFM) and ferromagnetic (FM) Heisenberg model as well as the quantum XY-model.
The Euclidean time interval $[0,\beta]$, where $\beta=1/T$ is the inverse temperature, together with the real-time interval $[0,t_N]$ then forms a closed Konstantinov-Perel' contour in the complex time plane. 
We again emphasize that $H$ is only used to prepare an ensemble of initial states whereas the real-time evolution is entirely driven by the dissipative measurement process.
In order to simulate this process, we will employ a multi-cluster algorithm \cite{Evertz:1992rb,Wiese:1994}.
We summarize the rules for forming loop-clusters for the different models and measurement processes in Appendix~\ref{app:cluster_rules}.

\section{Results and discussion}

We now come to the results regarding the dissipative measurement-driven dynamics of spin models, which are based on numerical simulations employing a loop-cluster algorithm.
We study in detail the dynamics for different measurement processes as summarized in Table~\ref{tab:processes} and show the crucial role of the symmetry of the measurement process for the real-time evolution.
In the following, we investigate the initial state corresponding to the antiferromagnetic Heisenberg model in great detail.
The initial states corresponding to the ferromagnetic Heisenberg model as well as to the quantum XY-model are discussed in Appendix~\ref{app:results_fm} and Appendix~\ref{app:results_xy}, respectively.

\begin{table}[b]
\begin{tabular}{|C{6cm}|C{3.5cm}|C{3.5cm}|C{3.5cm}|} \hline
           &$O^{(1)}$&$O^{(2)}$&$O^{(3)}$ \\ \hline
measured observable & $\vec{S}^2=(\vec{S}_x+\vec{S}_y)^2$&$S_x^{1}S_y^1$ or $S_x^{2}S_y^2$&$S_x^1S_y^1-S_x^2S_y^2$ \\ \hline
conserved Fourier mode $S(p)$ & $(0,0)$ & \textemdash & $(\pi,\pi)$ \\ \hline
final equilibrium values $A(p)$ & Eq.~\eqref{eq:final_tab_total} & Eq.~\eqref{eq:final_x} & Eq.~\eqref{eq:final_tab_rl} \\\hline
\end{tabular}
\caption{The measurement processes under consideration, along with their conserved Fourier mode and their final equilibrium values $A(p)$.}
\label{tab:processes}
\end{table}

\subsection{Heisenberg anti-ferromagnet initial state}
\label{sec:results_afm}

In this section, we consider an initial density matrix $\rho_0$ corresponding to the anti-ferromagnetic Heisenberg model \eqref{eq:heisenberg}, where we choose the $3$-direction as the quantization axis.
The ensemble of initial states is then prepared by means of the Euclidean-time cluster rules \eqref{eq:cluster_afm}.
In the following, the $3$-component of the staggered magnetization order parameter is denoted by 
\begin{equation}
 M_s=\sum_{x}(-1)^{x_1+x_2}S_x^3 \ .
\end{equation}
In order to analyze the subsequent real-time dynamics, it is useful to introduce the Fourier modes for the two-dimensional square lattice of size $L\times L$ according to
\begin{equation}
 \label{eq:ft_modes}
 S(p)=\sum_{x}{\exp\left(ipx\right)S_x^3}=\sum_{x}{\exp\left(ip_1x_1+ip_2x_2\right)S_x^3} \ .
\end{equation}
Accordingly, the $3$-component of the staggered magnetization will be denoted as the $(\pi,\pi)$-mode whereas the $3$-component of the uniform magnetization
\begin{equation}
 M=\sum_{x}S_x^3 
\end{equation}
is denoted as the $(0,0)$-mode.

\subsubsection{Dynamics for different measurement processes}

Starting with an initial ensemble at low temperature $\beta J=5L/2a=40$ on the square lattice of size $L\times L$ with $L=16a$, where $a$ is the lattice spacing, we investigate the dissipation-driven dynamics for different measurement processes.
We emphasize that the various measurement processes strongly differ in their conservation properties.
On the one hand, the measurement process $O^{(1)}$ conserves the $(0,0)$-mode whereas the measurement process $O^{(3)}$ results in the conservation of the $(\pi,\pi)$-mode.
On the other hand, the measurement process $O^{(2)}$ does not conserve any Fourier component.

\begin{figure}[t]
 \includegraphics[width=\columnwidth]{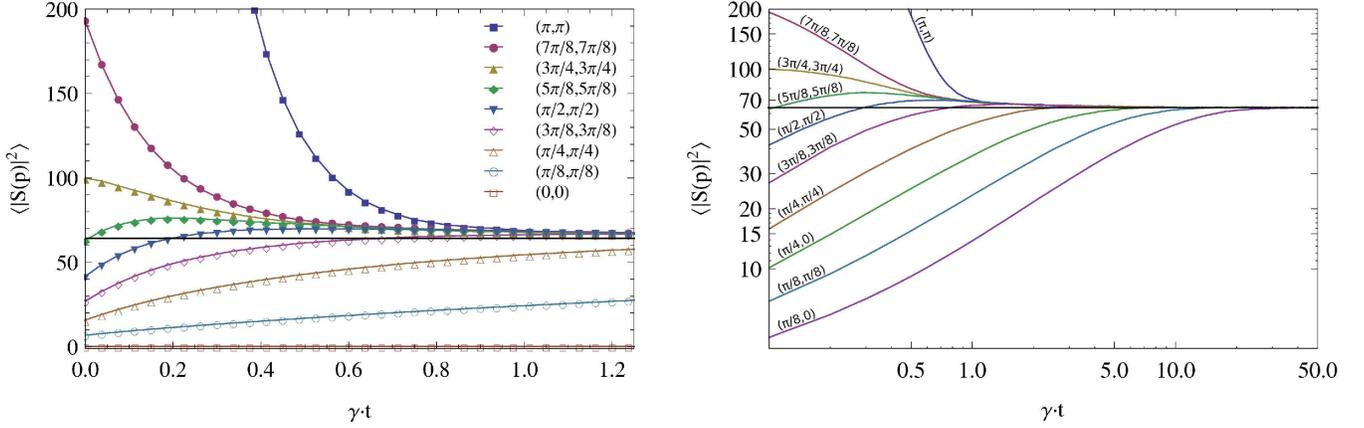}
 \caption{\label{fig:evolution1} 
 [Color online] (Heisenberg anti-ferromagnet initial state) 
 Time evolution of certain Fourier modes $\langle|S(p)|^2\rangle(t)$ for the measurement process $O^{(1)}$.
 {\it Left}: Linear plot for a short time interval, cf.~also Fig.~1c in \cite{Banerjee:2014yea}.
 {\it Right}: Log-log plot for a long time interval.
 The error bars are of the order of the symbol sizes and the lines are included to guide the eye.
 We initialize an anti-ferromagnetic Heisenberg model at low temperatures such that the $(\pi,\pi)$-mode is large whereas the $(0,0)$-mode vanishes.
 The remaining parameters are $4N_\tau=512$, $L=16a$, $\beta J = 5L/2a=40$ and $\epsilon\gamma=0.05$.
 In order to determine the time evolution, we performed $10^{6}$ Monte Carlo measurements.
 The horizontal line corresponds to the analytically derived final equilibrium value \eqref{eq:final_total}.
 We emphasize that the $(0,0)$-mode is exactly conserved during the time evolution.}
\end{figure}

\begin{figure}[t]
 \includegraphics[width=0.5\columnwidth]{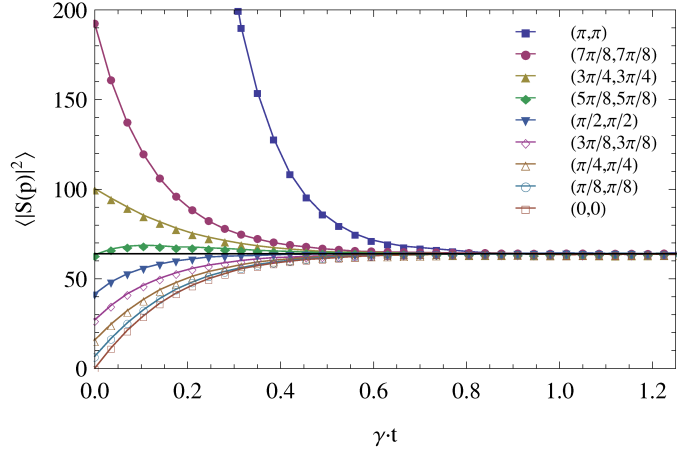}
 \caption{\label{fig:evolution2} 
 [Color online] (Heisenberg anti-ferromagnet initial state) 
 Time evolution of certain Fourier modes $\langle|S(p)|^2\rangle(t)$ for the measurement process $O^{(2)}$ on a linear plot.
 The error bars are again of the order of the symbol sizes and the lines are included to guide the eye.
 The parameters are as in Fig.~\ref{fig:evolution1} and the horizontal line corresponds to the analytically derived final equilibrium value \eqref{eq:final_x}.
 We emphasize that all Fourier modes equilibrate very rapidly.}
\end{figure}

\begin{figure}[t]
 \includegraphics[width=\columnwidth]{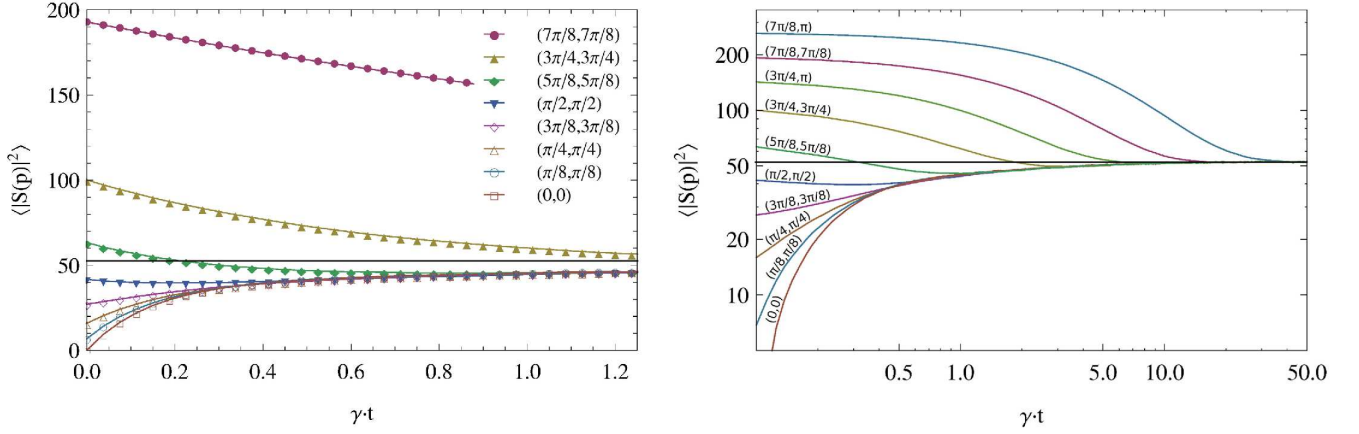}
 \caption{\label{fig:evolution3} 
 [Color online] (Heisenberg anti-ferromagnet initial state) 
 Time evolution of certain Fourier modes $\langle|S(p)|^2\rangle(t)$ for the measurement process $O^{(3)}$.
 {\it Left}: Linear plot for a short time interval.
 {\it Right}: Log-log plot for a long time interval.
 The error bars are again of the order of the symbol sizes and the lines are included to guide the eye.
 The parameters are as in Fig.~\ref{fig:evolution1} and the horizontal line corresponds to the analytically derived final equilibrium value \eqref{eq:final_rl}.
 We emphasize that the $(\pi,\pi)$-mode is exactly conserved during the time evolution.}
\end{figure}

Considering the Lindblad evolution with $\epsilon\gamma=0.05$ for different measurement processes, we display the real-time dynamics of a variety of Fourier modes
\begin{equation}
 \langle|S(p)|^2\rangle(t_k)=\tr{\rho(t_k)|S(p)|^2} \ ,
\end{equation}
with $k\in\{0,...,N\}$ in Figs.~\ref{fig:evolution1} -- \ref{fig:evolution3}.
We checked the dependence on varying $\epsilon\gamma$ to guarantee that we are effectively simulating a continuous Lindblad process with the discrete-time algorithm.
Most notably, the different conservation properties of the measurement processes are reflected in the time-dependence of the Fourier modes.
While the conserved quantities -- i.e. the $(0,0)$-mode for $O^{(1)}$ and the $(\pi,\pi)$-mode for $O^{(3)}$ -- do not equilibrate at all, the Fourier modes in the vicinity of the conserved quantity show a much slower equilibration rate than Fourier modes remote from the conserved quantity.
On the other hand, all Fourier modes show a rapid equilibration for the measurement process $O^{(2)}$ for which there are no conserved Fourier modes present.

After an initial phase, which is studied in more detail in the next section, the various Fourier modes approach their ultimate new equilibrium exponentially 
\begin{equation}
 \label{eq:equilibration}
 \langle|S(p)|^2\rangle(t) \stackrel{t\to\infty}{\simeq} A(p)+B(p)\exp\left(-t/\tau(p)\right) \ ,
\end{equation}
where $1/\tau(p)$ denotes the equilibration rate and $A(p)$ is the final equilibrium value.
Employing the definition \eqref{eq:ft_modes}, we find
\begin{equation}
 \label{eq:fourier_modes}
 \langle|S(p)|^2\rangle(t)=\sum_{x}\sum_{y}\exp\left(ip(x-y)\right)\langle S_{x}^3S_{y}^3\rangle(t) \ ,
\end{equation}
so that
\begin{equation}
 \label{eq:fourier_sum}
 \sum_{p}\langle|S(p)|^2\rangle(t)=N^2\sum_{x}\langle(S_{x}^{3})^2\rangle(t)=\frac{N^4}{4} \ ,
\end{equation}
where $N$ denotes the number of lattice points in each spatial direction such that $L=Na$.
Accordingly, in order to find $A(p)$, we need to calculate the spin-spin correlation function $\langle S_{x}^3S_{y}^3\rangle$ in the final state.
In fact, the final density matrix $\rho(t\to\infty)$, to which the system is driven by the measurement process, is constrained by the conserved quantities of the measurement process and is proportional to the unit matrix in each symmetry sector.

In Fig.~\ref{fig:evolution1}, we show the real-time evolution of a variety of Fourier modes for the measurement process $O^{(1)}$.
As already discussed, we find that the $(0,0)$-mode is exactly conserved during the time evolution.
On the other hand, the Fourier mode which is closest to the $(0,0)$-mode has the slowest equilibration rate whereas the Fourier modes which are further away equilibrate more rapidly.
As a matter of fact, all Fourier modes except the $(0,0)$-mode are then driven towards the same equilibrium value $A^{(1)}(p\neq(0,0))$, which will be discussed below in more detail.

The detailed equilibration, however, is rather intricate since a non-trivial attractor $\mathcal{A}(t)$ is formed which approaches $A^{(1)}(p\neq(0,0))$ for $t\to\infty$:
Due to the fact that the slowest Fourier mode approaches $A^{(1)}(p\neq(0,0))$ from below, all other modes finally reach it from above due to \eqref{eq:fourier_sum}.
This, however, has far reaching consequences for the time evolution of the single Fourier modes.
On the one hand, Fourier modes which start off above the attractor $\mathcal{A}(t)$ fall onto it from above so that their amplitudes decrease monotonically in time.
On the other hand, Fourier modes which start off below the attractor $\mathcal{A}(t)$ are enhanced in a first step so that they are driven towards the attractor from below.
Once these Fourier modes reach the attractor, their amplitude again evolves along the attractor and decreases monotonically in time.
As a consequence, all Fourier modes except the slowest one will at some point in time fall onto the attractor, which always lies above $A^{(1)}(p\neq(0,0))$.

Moreover, it turns out that the time needed to approach the attractor $\mathcal{A}(t)$ in the first instance depends on the specific Fourier mode.
As time evolves, more and more Fourier modes approach the attractor and, subsequently, show the same relaxation dynamics.
In the end, all Fourier modes except the slowest one lie on the attractor such that the final approach towards the new equilibrium at $t\to\infty$ is determined by the equilibration rate $1/\tau(p)$ of the slowest Fourier mode.

In order to derive an analytic expression for $A^{(1)}(p\neq(0,0))$, we consider the system to be prepared at low temperature $T\to0$ at the initial time $t_0$, such that
\begin{equation}
 A^{(1)}(p=(0,0))=\langle|S(p=(0,0))|^2\rangle(t_0)\simeq0 \ ,
\end{equation}
corresponding to the ground state of the system for which the uniform magnetization vanishes.
Due to the fact that the $(0,0)$-mode is exactly conserved during the time evolution, the system is driven to a final equilibrium ensemble for which the $3$-component of the uniform magnetization still vanishes.
In general, the final state partition function in any sector of magnetization $M\in\{-N^2/2,...,N^2/2\}$ is calculated according to
\begin{equation}
 Z_M[j]=\prod_{x}\sum_{S_x^{3}=\pm\frac{1}{2}}\exp\Big(\sum_z{j_zS^3_{z}}\Big)\delta_{\sum_z{S^3_{z}},M}=
 \frac{1}{2\pi}\int\limits_{0}^{2\pi}{d\lambda}\prod_{x}\sum_{S_x^{3}=\pm\frac{1}{2}}\exp\Big(\sum_z{(i\lambda+j_z)S^3_{z}}-i\lambda M\Big) \ .
\end{equation}
Restricting ourselves to the $M=0$ sector of the Hilbert space according to the chosen initial state, we find that there are
\begin{equation}
 Z_{M=0}[0]=\frac{1}{2\pi}\int\limits_{0}^{2\pi}d\lambda\left[2\cos\left(\tfrac{\lambda}{2}\right)\right]^{N^2}=\begin{pmatrix}N^2\\\frac{N^2}{2}\end{pmatrix}=\frac{N^2!}{\frac{N^2}{2}!\frac{N^2}{2}!}
\end{equation}
states which are equally probable.
The spin-spin-correlation function is then calculated according to
\begin{equation}
 \langle S_{x}^3S_{y}^3\rangle=
  \frac{1}{2\pi Z_{M=0}[0]}\int\limits_{0}^{2\pi}{d\lambda}\prod_{x}\sum_{S_x^{3}=\pm\frac{1}{2}}S^3_{x}S^3_{y}\exp\Big(\sum_z{i\lambda S^3_{z}}\Big) \ .
\end{equation}
We trivially obtain $\tfrac{1}{4}$ for $x=y$.
On the other hand, for $x\neq y$ we find
\begin{equation}
 \langle S_{x}^3S_{y}^3\rangle=
 -\frac{1}{8\pi Z_{M=0}[0]}\int\limits_{0}^{2\pi}d\lambda\left[2\cos\left(\tfrac{\lambda}{2}\right)\right]^{N^2}\tan^2\left(\tfrac{\lambda}{2}\right)=-\frac{1}{4(N^2-1)} \ .
\end{equation}
Accordingly, the spin-spin-correlation function can be written as
\begin{equation}
 \langle S_{x}^3S_{y}^3\rangle=\frac{N^2\delta_{x,y}-1}{4(N^2-1)} \ .
\end{equation}
As a consequence, upon evaluating \eqref{eq:fourier_modes} we find that the Fourier modes take the final values
\begin{subequations}
\label{eq:final_tab_total}
\begin{align}
 A^{(1)}(p=(0,0))&=0 \ , \\
 A^{(1)}(p\neq(0,0))&=\frac{N^4}{4(N^2-1)} \ .
 \label{eq:final_total}
\end{align}
\end{subequations}

The measurement process $O^{(2)}$, on the other hand, has the simplest final state as it does not conserve any Fourier mode.
As a consequence, all Fourier modes equilibrate very quickly as shown in Fig.~\ref{fig:evolution2}.
We note that the equilibration of the distinct Fourier modes is not completely independent from each other because of \eqref{eq:fourier_sum}, which gives rise to an overshooting of several Fourier modes, e.~g. the $(5\pi/8,5\pi/8)$-mode in the figure.
However, we do not observe the occurrence of a non-trivial attractor as there is no slow mode present in the system.
In fact, the final equilibrium corresponds to a state where each of the $2^{N^2}$ spin states is equally probable.
Therefore, the spin-spin-correlation function in the final equilibrium is easily calculated according to
\begin{equation}
 \langle S_{x}^3S_{y}^3\rangle=\frac{1}{2^{N^2}}\prod_{x}\sum_{S^3_{x}=\pm\frac{1}{2}}S^3_{x}S^{3}_{y}=\frac{1}{4}\delta_{x,y} \ ,
\end{equation}
so that all Fourier modes take the same final value
\begin{equation}
 \label{eq:final_x}
 A^{(2)}(p)=\frac{N^2}{4} \ .
\end{equation}
 
Finally, we display the real-time evolution of the Fourier modes for the measurement process $O^{(3)}$ in Fig.~\ref{fig:evolution3}.
As already discussed, this measurement process conserves the $(\pi,\pi)$-mode, corresponding to the order parameter of the system which is large at low temperatures $T\to0$.
In fact, finite-volume chiral perturbation theory predicts
\begin{equation}
 \label{eq:analytic_order}
 A^{(3)}(p=(\pi,\pi))=\langle|S(p=(\pi,\pi)|^2\rangle(t_0)\simeq\frac{\mathcal{M}_s^2 L^4}{3}\sum_{n} a_n\left(\frac{c}{\rho_sL}\right)^n  \ ,
\end{equation}
where the first three coefficients are given by $a_0=1$, $a_1=0.45157$ and $a_2=0.082803$ \cite{Gockeler:1990ik,Gockeler:1990zn,Hasenfratz:1993}.
The low-energy parameters are the staggered magnetization density $\mathcal{M}_s=0.30743(1)/a^2$, the spin stiffness $\rho_s=0.1808(4)J$ and the spin-wave velocity $c=1.6585(10)Ja$ \cite{Sandvik:2010,Gerber:2009rd}.
Accordingly, the typical length scale is given by $\xi=c/(2\pi\rho_s)=1.459(3)a$.

Due to the fact that both measurement processes $O^{(1)}$ and $O^{(3)}$ have a conserved mode, the real-time evolution of the Fourier modes shows several similarities, most notably the occurrence of a non-trivial attractor $\mathcal{A}(t)$.
Nevertheless, there is also an important difference:
The Fourier mode with the slowest equilibration rate towards the final equilibrium value $A^{(3)}(p\neq(\pi,\pi))$ is the one which lies closest to the $(\pi,\pi)$-mode. 
As a consequence, the slowest Fourier mode approaches $A^{(3)}(p\neq(\pi,\pi))$ from above whereas all other modes finally reach it from below due to \eqref{eq:fourier_sum}.
Accordingly, the corresponding attractor $\mathcal{A}(t)$ lies below $A^{(3)}(p\neq(\pi,\pi))$ and increases monotonically in time.
It has to be emphasized, however, that the attractor's distance from the final equilibrium value $A^{(3)}(p\neq(\pi,\pi))$ at early times is much larger than for the measurement process $O^{(1)}$ due to the larger amplitude of the slowest mode.
In similarity to the measurement process $O^{(1)}$, however, all Fourier modes except the slowest one will, at some point in time, fall onto the attractor such that their final approach towards equilibrium is determined by the equilibration rate of the slowest mode.

In order to derive a simple approximation for the final equilibrium value $A^{(3)}(p\neq(\pi,\pi))$, we construct the final state partition function in any sector of staggered magnetization $M_s\in\{-N^2/2,...,N^2/2\}$ according to
\begin{equation}
 Z_{M_s}[j]=\prod_{x}\sum_{S_x^{3}=\pm\frac{1}{2}}\exp\Big(\sum_z{j_zS^3_{z}}\Big)\delta_{\sum_z{(-1)^{z_1+z_2}S^3_{z}},M_s}=
 \frac{1}{2\pi}{\displaystyle\int\limits_{0}^{2\pi}}{d\lambda}\prod_{x}\sum_{S_x^{3}=\pm\frac{1}{2}}\exp\Big(\sum_z{(i\lambda(-1)^{z_1+z_2}+j_z)S^3_{z}}-i\lambda M_s\Big) \, .
\end{equation}
Within each of the $M_s$-sectors, we find that there are
\begin{equation}
 Z_{M_s}[0]=\frac{1}{2\pi}\int\limits_{0}^{2\pi}d\lambda\left[2\cos\left(\tfrac{\lambda}{2}\right)\right]^{N^2}\exp\left(-i\lambda M_s\right)=\begin{pmatrix}N^2\\\frac{N^2}{2}+M_s\end{pmatrix}=\frac{N^2!}{(\frac{N^2}{2}+M_s)!(\frac{N^2}{2}-M_s)!}
\end{equation}
equally probable states. 
In fact, we can again calculate the spin-spin-correlation function in any $M_s$-sector via
\begin{equation}
 \langle S_{x}^3S_{y}^3\rangle_{M_s}=
 \frac{1}{2\pi Z_{M_s}[0]}\int\limits_{0}^{2\pi}{d\lambda}\prod_{x}\sum_{S_x^{3}=\pm\frac{1}{2}}S^3_{x}S^3_{y}\exp\Big(\sum_z{i\lambda(-1)^{z_1+z_2} S^3_{z}}-i\lambda M_s\Big) \ .
\end{equation}
Again, for $x=y$ we trivially obtain $\tfrac{1}{4}$, whereas in all other cases
\begin{equation}
 \langle S_{x}^3S_{y}^3\rangle_{M_s}=
 -\frac{\varphi_{x,y}}{8\pi Z_{M_s}[0]}\int\limits_{0}^{2\pi}d\lambda\left[2\cos\left(\tfrac{\lambda}{2}\right)\right]^{N^2}\tan^2\left(\tfrac{\lambda}{2}\right)\exp\left(-i\lambda M_s\right)=
 -\varphi_{x,y}\frac{N^2-(2M_s)^2}{4N^2(N^2-1)} \ ,
\end{equation}
with $\varphi_{x,y}=(-1)^{x_{1}+x_{2}+y_{1}+y_{2}}=\pm1$, depending on whether $x$ and $y$ are on the same staggered sublattice or not.
Accordingly, the spin-spin-correlation function in any $M_s$-sector can be written as
\begin{equation}
 \langle S_{x}^3S_{y}^3\rangle_{M_s}=\frac{[N^4-(2M_s)^2]\delta_{x,y}-[N^2-(2M_s)^2]\varphi_{x,y}}{4N^2(N^2-1)} \ .
\end{equation}
It has been shown \cite{Gerber:2009rd} that the probability $p(M_s)$ of being in a certain $M_s$-sector at $t_0$ is well-approximated by a discrete uniform distribution:
\begin{equation}
 p(M_s)=\begin{cases}\frac{1}{2M_{s,\max}+1}&\quad |M_s|\leq M_{s,\max}\\0&\quad|M_s|>M_{s,\max} \end{cases}
\end{equation}
In fact, $M_{s,\max}=\lceil \mathcal{M}_sL^2\rceil=\lceil0.30743 N^2\rceil$ in the infinite volume limit $N\to\infty$ at $T\to0$.
Here, $\lceil X \rceil$ denotes the ceiling function which returns the smallest integer not less than $X$.
For finite values of $N$, however, there are deviations so that $M_{s,\max}>\lceil0.30743 N^2\rceil$, e.~g. $M_{s,\max}\simeq\lceil0.39N^2\rceil=25$ for $N=8$, $M_{s,\max}\simeq\lceil0.345N^2\rceil=89$ for $N=16$ and $M_{s,\max}\simeq\lceil0.335N^2\rceil=192$ for $N=24$ \cite{Gerber:2009rd}. 
As the measurement-driven real-time evolution does not change the $M_s$-sector of any ensemble member, this distribution $p(M_s)$ does not change with time.
Accordingly, by averaging over all allowed $M_s$-sectors and evaluating \eqref{eq:fourier_modes} we find for the Fourier modes
\begin{subequations}
\label{eq:final_tab_rl}
\begin{align}
 A^{(3)}(p=(\pi,\pi))&=\langle|S(p=(\pi,\pi))|^2\rangle(t_0)\simeq\frac{1}{3}M_{s,\max}(M_{s,\max}+1) \ , \\
 \label{eq:final_rl}
 A^{(3)}(p\neq(\pi,\pi))&=\frac{1}{N^2-1}\left[\frac{N^4}{4}-\langle|S(p=(\pi,\pi))|^2\rangle(t_0)\right]\simeq\frac{1}{N^2-1}\left[\frac{N^4}{4}-\frac{1}{3}M_{s,\max}(M_{s,\max}+1)\right] \ .
\end{align}
\end{subequations}
We emphasize that the relative error of this approximation with respect to the analytic value \eqref{eq:analytic_order} is only of the order of a few percent and vanishes in the infinite volume limit $N\to\infty$.

\subsubsection{Equilibration and attraction time scales}

We have seen that the behavior of the Fourier modes strongly depends on the existence of conserved quantities in the measurement process as well as on the momentum value $p$. 
In the following, we concentrate on the most interesting cases of the measurement processes $O^{(1)}$ and $O^{(3)}$.
In these cases, we observed the existence of a non-trivial attractor $\mathcal{A}(t)$ towards which all equilibrating Fourier modes except the slowest one are driven before the final equilibrium value $A(p)$ is reached.
As a consequence, the approach towards the final equilibrium is slowed down by the slowest Fourier mode in the system.
Moreover, larger systems are supposed to equilibrate ever slower due to the presence of Fourier modes with ever smaller momentum.

In the following, we further investigate the equilibration dynamics for measurement process $O^{(1)}$.
We emphasize that an analogous discussion also applies to measurement process $O^{(3)}$.
First, we study the final approach towards equilibrium.
To this end, suppose that all Fourier modes except the slowest one (note that there are in total four of them on the isotropic square lattice) with momentum value
\begin{equation}
 \label{eq:momentum}
 |p_1|=\frac{2\pi}{aN}
\end{equation}
have already fallen onto the attractor.
In fact, the time-dependence of the slowest Fourier mode is well-described by an exponential decay
\begin{equation}
  \langle|S(p_1)|^2\rangle(t) =A+C(p_1)\exp\left(-\frac{t}{\tau(p_1)}\right) \ , 
\end{equation}
with $A=N^4/4(N^2-1)$.
Having measured the parameters $C(p_1)<0$ and $\tau(p_1)$ and taking into account the sum rule \eqref{eq:fourier_sum}, we can determine the ultimate approach of the attractor towards the final equilibrium value according to
\begin{equation}
 \langle|S(p\neq p_1)|^2\rangle(t) \stackrel{t>t_2}{\simeq} A+B(p)\exp\left(-\frac{t}{\tau(p)}\right)=A - \frac{4C(p_1)}{N^2-5}\exp\left(-\frac{t}{\tau(p_1)}\right) \ ,
\end{equation}
where $t_2$ denotes the time at which the second-slowest Fourier mode with momentum $p_2$ has fallen onto the attractor.
This clearly shows that the final equilibration rate $1/\tau(p)$ of all Fourier modes is determined by the behavior of the slowest Fourier mode
\begin{align}
 \tau(p)&=\tau(p_1) \ .
\end{align}

In fact, the second-slowest Fourier mode (again, there are in total four modes with the same value of $|p_2|$) is then well-described by a sum of two exponential functions
\begin{equation}
 \langle|S(p_2)|^2\rangle(t) = A-\frac{4C(p_1)}{N^2-5}\exp\left(-\frac{t}{\tau(p_1)}\right)+C(p_2)\exp\left(-\frac{t}{T(p_2)}\right) \ , 
\end{equation}
one of them describing the approach towards the attractor and the other the flow along the attractor.
We can again numerically determine the coefficient $C(p_2)<0$ and the relaxation time $1/T(p_2)$.
Employing the sum rule \eqref{eq:fourier_sum} again, we can deduce the attractor according to
\begin{equation}
 \langle|S(p\neq p_{1,2})|^2\rangle(t) \stackrel{t>t_3}{\simeq} A - \frac{4C(p_1)}{N^2-5}\exp\left(-\frac{t}{\tau(p_1)}\right)-\frac{4C(p_2)}{N^2-9}\exp\left(-\frac{t}{T(p_2)}\right) \ ,
\end{equation}
where $t_3<t_2$ denotes the time at which the third-slowest Fourier mode with momentum $p_3$ has fallen onto the attractor.

\begin{figure}[t]
 \includegraphics[width=\columnwidth]{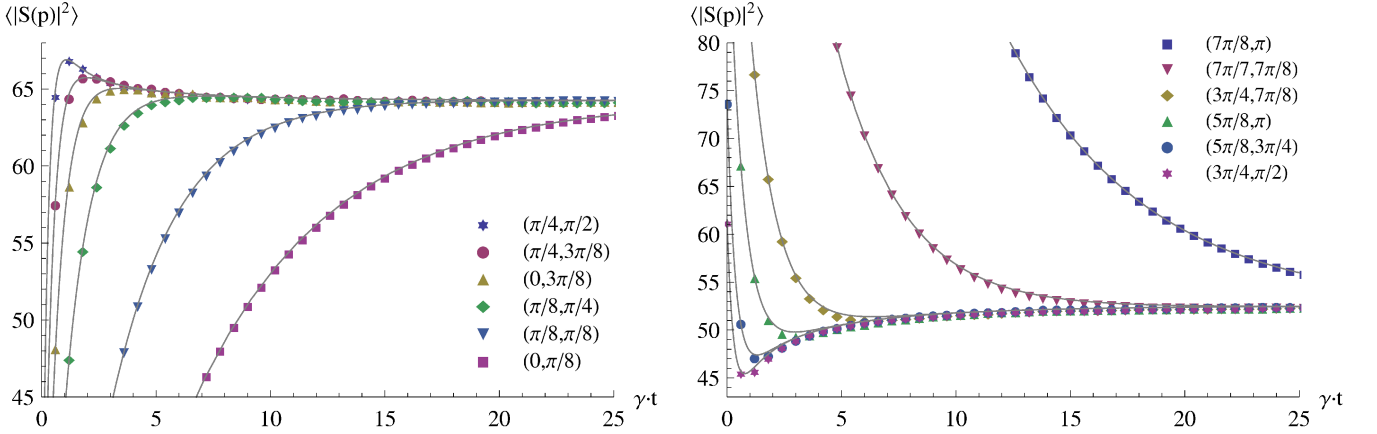}
 \caption{\label{fig:attractor} 
 [Color online] (Heisenberg anti-ferromagnet initial state) 
 Time evolution of certain Fourier modes $\langle|S(p)|^2\rangle(t)$ for the measurement processes $O^{(1)}$ ({\it left}) and $O^{(3)}$ ({\it right}).
 The error bars are of the order of the symbol sizes.
 The solid lines correspond to a simultaneous fit of the twelve slowest Fourier modes upon taking into account the analytically predicted attractor behavior.
 The parameters are $4N_\tau=512$, $L=16a$, $\beta J = 5L/2a=40$, $\epsilon\gamma=0.05$ and we performed $4\cdot10^6$ Monte Carlo measurements.}
\end{figure}

In principle, this procedure can be continued for all momentum modes and therefore allows for the iterative determination of the attractor
\begin{equation}
 \label{eq:attractor}
 \mathcal{A}(t)=\frac{N^4}{4(N^2-1)}+\sum_{p}{\widetilde{C}(p)\exp\left(-\frac{t}{T(p)}\right)} \ ,
\end{equation}
with $T(p_1)\equiv\tau(p_1)$.
Most important, the temporal dependence of the attractor is determined by a sum of exponential functions with different attraction time scales $T(p)$, which describe the approach of the single momentum modes towards the attractor.
It is clear, however, that this method becomes increasingly involved for larger momenta due to the numerical uncertainties in the data.
Moreover, for ever larger momenta the order in which the single Fourier modes approach the attractor becomes less clear.
For the lowest Fourier modes, however, all these issues are under control and we can reliably determine the parameters $C(p)$ and $T(p)$.

\begin{figure}[b]
 \includegraphics[width=\columnwidth]{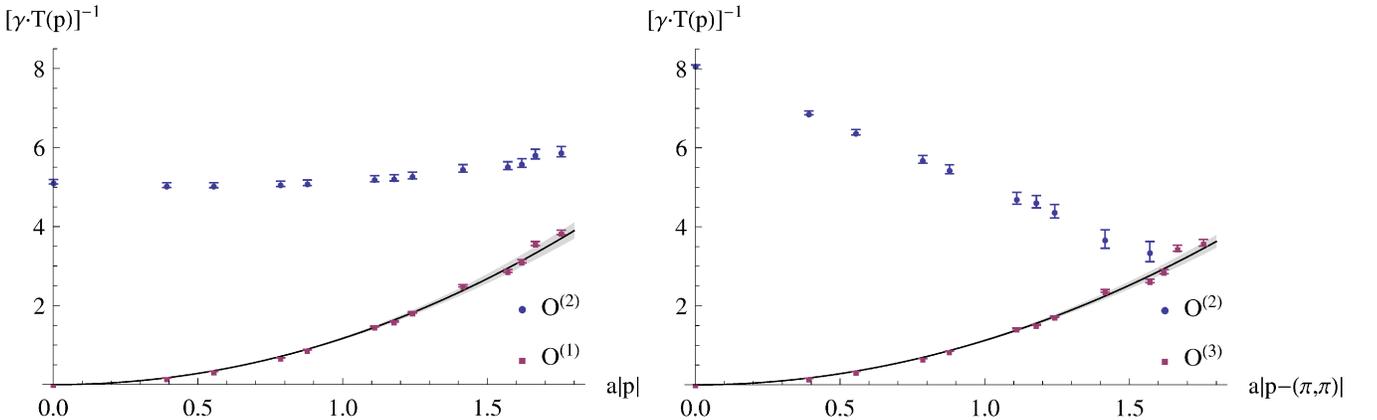}
 \caption{\label{fig:momentum_dependence} 
 [Color online] (Heisenberg anti-ferromagnet initial state) 
 Momentum dependence of the attraction rate $1/T(p)$ for the measurement processes $O^{(1)}$ and $O^{(2)}$ ({\it left}) as well as $O^{(3)}$ and $O^{(2)}$ ({\it right}).
 The data points for the measurement processes $O^{(1)}$ and $O^{(3)}$, respectively, are calculated by a simultaneous fit of the twelve slowest Fourier modes upon taking into account the analytically predicted attractor behavior.
 On the other hand, we performed independent exponential fits for the measurement process $O^{(2)}$. 
 The parameters are as in Fig.~\ref{fig:attractor}.}
\end{figure}

In fact, we can simultaneously fit the slowest Fourier modes -- i.e. the Fourier modes in the vicinity of the $(0,0)$-mode for $O^{(1)}$ and in the vicinity of the $(\pi,\pi)$-mode for $O^{(3)}$, respectively -- upon taking into account the analytically predicted attractor behavior.
We show the good agreement of these fits with the numerical data in Fig.~\ref{fig:attractor}. 
The precise determination of the parameters further allows us to study the momentum dependence of the time scale $T(p)$, which determines the rate at which the different Fourier modes are driven towards the attractor.
Fig.~\ref{fig:momentum_dependence} displays the momentum dependence of the attraction rates for the different measurement processes, along with the best fit of the corresponding data.
Most notably, for $O^{(1)}$ we find in the small momentum regime
\begin{equation}
 \label{eq:mom_dep_o1}
 [\gamma T(p)]^{-1}=c_{1}\left(a|p|\right)^{r_{1}} \ ,
\end{equation}
with $c_{1}=1.17(2)$ and $r_{1}=2.05(4)$, cf.~also Fig.~1b and the corresponding analysis in \cite{Banerjee:2014yea}.
On the other hand, for $O^{(3)}$ we obtain in the large momentum regime
\begin{equation}
 \label{eq:mom_dep_o3}
 [\gamma T(p)]^{-1}=c_{3}\left(a|p-(\pi,\pi)|\right)^{r_{3}} \ ,
\end{equation}
with $c_{3}=1.12(2)$ and $r_{3}=1.99(4)$. 
Obviously, the attraction rate shows for both measurement processes a quadratic momentum dependence.
Due to this behavior, we regard the approach towards the attractor as a diffusion process of the conserved quantity.
Larger systems are supposed to equilibrate ever slower due to the presence of Fourier modes with ever smaller momentum \eqref{eq:momentum}.
In the infinite volume limit $N\to\infty$, the momentum variable becomes continuous and the equilibration rate \eqref{eq:mom_dep_o1} and \eqref{eq:mom_dep_o3}, respectively, becomes arbitrarily small.

This is contrasted by the momentum dependence of the equilibration rate for the measurement process $O^{(2)}$ which does not conserve any of the Fourier modes.
In this case, all Fourier modes equilibrate very quickly as there is no additional constraint which could delay the approach towards the new equilibrium state.

\subsubsection{Non-equilibrium phase transition}

In the anti-ferromagnetic Heisenberg model, at low temperatures $T\to0$ according to \eqref{eq:analytic_order} we find the behavior
\begin{equation}
 \frac{\langle M_s^2\rangle(t_0)}{L^2}=\frac{\langle|S(p=(\pi,\pi)|^2\rangle(t_0)}{L^2}\sim L^2 \ , 
\end{equation}
indicating spontaneous symmetry breaking of the $SU(2)$ spin symmetry.
During the Lindblad evolution, this order is then destroyed by the measurement processes $O^{(1)}$ and $O^{(2)}$, respectively, whereas it is conserved for the measurement process $O^{(3)}$.
In the first two cases, the system is driven towards the final states \eqref{eq:final_tab_total} and \eqref{eq:final_x}, respectively, for which $\langle M_s^2\rangle/L^2$ becomes volume-independent, indicating that the $SU(2)$ spin symmetry is restored again.
In the infinite volume limit, the system evolves from an ordered state with a finite order parameter density to a disordered state where the order parameter density vanishes.
Accordingly, the system must undergo a phase transition at some point, however, it is not clear a-priori whether this transition occurs at a certain instant of time or rather takes an infinite amount of time.
In fact, since the Lindblad evolution drives the system far out-of-equilibrium, this phase transition is not expected to fall into any of the standard dynamical universality classes \cite{Hohenberg:1977ym}.

\begin{figure}[t]
 \includegraphics[width=0.5\columnwidth]{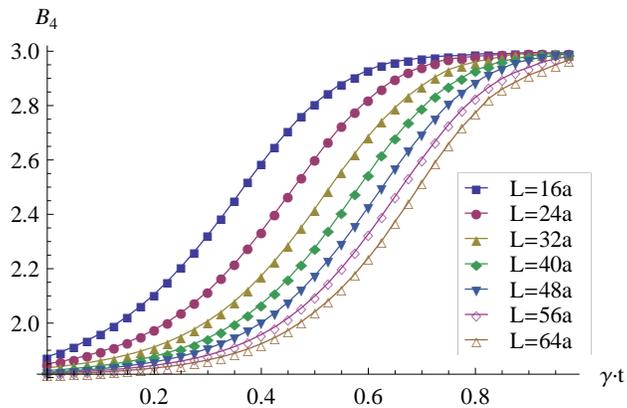}
 \caption{\label{fig:binder_ratio} 
 [Color online] (Heisenberg anti-ferromagnet initial state) 
 Time-dependent Binder ratio $B_4(t)$ for the measurement process $O^{(2)}$ for different values of $L/a$ along with variable $\beta J=3L/4a$.
 The error bars are of the order of the symbol sizes and the lines are included to guide the eye.
 We emphasize that the results for the measurement process $O^{(1)}$ are essentially identical, cf.~also Fig.~2b in \cite{Banerjee:2014yea}.}
\end{figure}

In order to study the temporal behavior of the phase transition, we display the real-time evolution of the Binder ratio \cite{Binder:1981zz,Binder:1984}
\begin{equation}
 B_4(t)=\frac{\langle M_s^4\rangle(t)}{[\langle M_s^2\rangle(t)]^2} 
\end{equation}
in Fig.~\ref{fig:binder_ratio}.
Typically, for systems in thermal equilibrium plotting $B_4$ versus the temperature for different system sizes produces curves that intersect each other close to the phase transition temperature.
Doing the same thing for the real-time evolution, we observe that the various finite-volume curves for $B_4(t)$ do not intersect each other. 
Moreover, we find that their inflection points move to ever later times with increasing volumes.
We interpret this result as saying that the phase transition does not occur at any specific instant in time but takes rather an infinite amount of time before it is completed.
It is quite remarkable that the results for the measurement processes $O^{(1)}$ and $O^{(2)}$ are essentially identical even though their conservation properties and loop-cluster rules are completely different.
This indicates that the destruction of the anti-ferromagnetic order is rather insensitive to the specific dissipative process.

\begin{figure}[t]
 \includegraphics[width=\columnwidth]{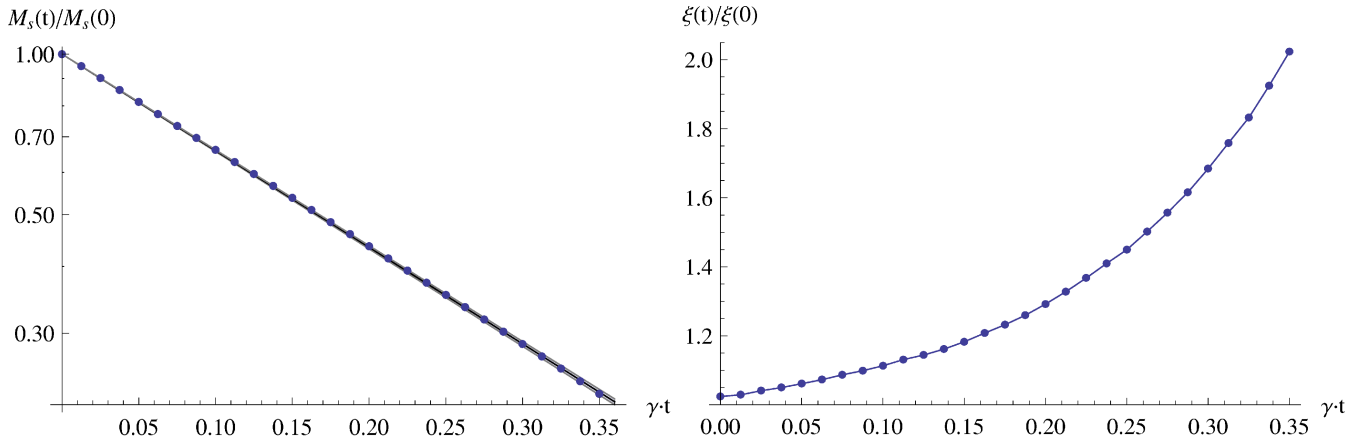}
 \caption{
 \label{fig:staggered_density}
 [Color online] (Heisenberg anti-ferromagnet initial state) 
 {\it Left:} Time-dependent staggered magnetization density $\mathcal{M}_s(t)/\mathcal{M}_s(0)$ for the measurement process $O^{(2)}$ on a logarithmic plot along with an exponential fit of the data.
 {\it Right:} Time-dependent length scale $\xi(t)/\xi(0)$ for the measurement process $O^{(2)}$ on a linear plot with the lines included to guide the eye.
 In both cases, the error bars are of the order of the symbol sizes.
 We again emphasize that the results for the measurement process $O^{(1)}$ are essentially identical, cf.~also Figs.~2c-d in \cite{Banerjee:2014yea}.}
\end{figure}

In order to further study the dynamics of the phase transition, we investigated the time-dependence of the staggered magnetization density $\mathcal{M}_s(t)$.
Based on finite-volume chiral perturbation theory \eqref{eq:analytic_order}, we first performed a finite size analysis at $t_0$ to determine the shape coefficients $c_n$ up to forth order according to
\begin{equation}
 \langle M_s^2\rangle(t_0) = \frac{\mathcal{M}_s^2(t_0) L^4}{3} \sum_{n=0}^{3}{c_n\left(\frac{\xi(t_0)}{L}\right)^n} \ .
\end{equation} 
As a reminder, the best numerical values for the used low-energy parameters are $\mathcal{M}_s(t_0)=0.30743(1)/a^2$ and $\xi(t_0)=1.459(3)a$ \cite{Sandvik:2010,Gerber:2009rd}.
We used finite-volume data up to $L=64a$ as well as results from \cite{Sandvik:2010} to obtain $c_0=1$, $c_1=5.7503(6)$, $c_2=16.31(2)$ and $c_3=-84.8(2)$.
The shape coefficients $c_n$, which depend on the geometry of the quantum system, are then assumed to be time-independent as the spatial geometry remains unchanged.
Moreover, we suppose that the chiral-perturbation-theory-inspired formula remains valid in real-time, so that
\begin{equation}
 \langle M_s^2\rangle(t) = \frac{\mathcal{M}_s^2(t) L^4}{3} \sum_{n=0}^{3}{c_n\left(\frac{\xi(t)}{L}\right)^n}=\frac{\mathcal{M}_s^2(t) L^4}{3} \sum_{n=0}^{3}{c_n\left(\frac{\xi(0)}{L}\right)^n\left(\frac{\xi(t)}{\xi(0)}\right)^n} \ .
\end{equation}
This then allows us to study the decay of the staggered magnetization in the infinite volume limit up to $\gamma t \simeq 0.35$ with an acceptable $\chi^2/\text{d.o.f.}$.
The decay rates can then be estimated from an exponential fit
\begin{equation}
 \mathcal{M}_s(t)=\mathcal{M}_s(0)\exp\left(-t/\tau \right) \ ,
\end{equation}
with an inverse decay rate $\gamma\tau=0.240(2)$, as shown in Fig.~\ref{fig:staggered_density}.
Again, this suggests that the phase transition is completed only after an infinite amount of time rather than after a finite time interval.
Additionally, we find that the length scale $\xi(t)/\xi(0)$ increases with time, which can be attributed to a decrease of the spin stiffness $\rho_s$.

\section{Conclusions \& outlook}
\label{sec:conclusion}

We investigated the real-time evolution of large strongly-coupled $2$-dimensional systems of quantum spins $s=\tfrac{1}{2}$ whose dynamics are entirely driven by measurements of neighboring spin pairs.
We considered different thermal initial states at low temperature corresponding to the anti-ferromagnetic or ferromagnetic Heisenberg model as well as the XY-model.
The dissipative time evolution, described by the Kossakowski-Lindblad equation, destroys the long-range order of the initial states and drives the systems to a new equilibrium with only short-range correlations.
To deepen the understanding of the real-time dynamics, we studied different measurement processes, which differed from each other in their symmetry properties, and derived the corresponding loop-cluster rules.
The path integral along the Konstantinov-Perel' contour was then solved with an efficient loop-cluster algorithm.

We studied in detail the anti-ferromagnetic Heisenberg model initial state for three different measurement processes, of which two result in the conservation of the $(0,0)$-mode (uniform magnetization) or of the $(\pi,\pi)$-mode (staggered magnetization), respectively. 
The conservation of any of the Fourier modes leads to a drastic slowing down of the equilibration process due to the occurrence of a non-trivial attractor, towards which all equilibrating Fourier modes except the slowest one are driven before the final equilibrium value $A(p)$ is reached.
As a consequence, the approach towards the final equilibrium is determined by the slowest Fourier mode in the system.
In fact, a detailed study of the relaxation rate $1/T(p)$ at which the different Fourier modes are driven towards the attractor exhibits a quadratic momentum dependence.
As a consequence, larger systems are supposed to equilibrate ever slower due to the presence of Fourier modes with ever smaller momentum.
Due to the quadratic momentum dependence, we identify the approach towards the attractor as a diffusion process of the conserved quantities.

This behavior is in contrast to the case of a dissipative process which does not conserve any of the Fourier modes.
In this case, all Fourier modes equilibrate very quickly as there is no additional constraint (besides an ever-present sum rule) which could delay the approach towards the final equilibrium state.
As a consequence, we do neither observe slow momentum modes nor a non-trivial attractor towards which the Fourier modes are driven.
Qualitatively, the same behavior is also found for initial states corresponding to the ferromagnetic Heisenberg model and the XY-model.

Studying the Heisenberg model at low temperatures, we find that $\langle M_s^2\rangle/L^2$ (anti-ferromagnet) or $\langle M^2\rangle/L^2$ (ferromagnet) are proportional to $L^2$, indicating spontaneous symmetry breaking of the underlying $SU(2)$ spin symmetry at zero temperature.
Similarly, the large value of $\langle M^2\rangle/L^2\sim L^2$ in the ferromagnetic XY-model indicates the quasi long-range order below the Kosterlitz-Thouless transition temperature.
After driving the system with various dissipative processes to its new equilibrium state, we find a disordering of the systems corresponding to the restoration of the spin symmetry. 
For the cases under consideration, we find that the staggered magnetization density (anti-ferromagnetic initial state) or the uniform magnetization density (ferromagnetic initial state) shows an exponential decay in time.
This suggests that a non-equilibrium phase transition does not occur at any finite instant in time but is rather completed only after an infinite amount of time.

A drawback of the current study is the fact that the real-time dynamics is entirely driven by the measurement process whereas the influence of the unitary Hamiltonian evolution has been disregarded in order to enable Monte Carlo importance sampling.
Nevertheless, its study is still worthwhile as this idealized dynamics may become realizable in optical lattice experiments with ultracold atoms in the future.
In fact, the main obstacle for including the effect of the Hamiltonian so far has been the complex weight problem which arises from the unitarity of the time-evolution operator.
We are currently exploring whether the meron-cluster idea, which has already been successfully employed in solving severe sign problems \cite{Bietenholz:1995zk,Chandrasekharan:1999cm}, is also applicable for solving the Lindblad dynamics including a Hamiltonian.

However, even though we did not explicitly take into account the Hamiltonian in our numerical simulations, we can still make statements about the long-time behavior of the full Lindblad evolution, i.~e. the combined dissipative Markovian and unitary Hamiltonian dynamics.
Regarding the Heisenberg model, for instance, the unit density matrix corresponding to \eqref{eq:final_x} is a stable $T\to\infty$ fixed point for a large class of measurement processes, including $O^{(2)}$ and $O^{(3)}$.
For the measurement process $O^{(1)}$, however, the system is still supposed to be driven towards the fixed points corresponding to \eqref{eq:final_tab_total} or \eqref{eq:final_total_fm}, respectively.
In this respect, the inclusion of the Hamiltonian could only change the characteristic time scales of the problem but not the fixed point structure.


\subsection*{Acknowledgments}

We like to thank M.~Kon for his collaboration on \cite{Banerjee:2014yea}, which provides the foundation for the work presented here.
The research leading to these results has received funding from the Ministry of Science and Technology (MOST) of Taiwan under grant number 102-2112-M-003-004-MY3, from the Schweizerische Nationalfonds and from the European Research Council under the European Union's Seventh Framework Programme (FP7/2007-2013)/ ERC grant agreement 339220. 


\appendix
\section{Loop cluster algorithm}
\label{app:cluster_rules}

In order to simulate the real-time dynamics, we employ an efficient multi-cluster algorithm \cite{Evertz:1992rb,Wiese:1994}.
In this algorithm, local stochastic decisions, which are determined by cluster rules, result in non-local changes of worldline configurations.
Compared to other Monte Carlo procedures, this allows for a drastic reduction of autocorrelation times.
Details on cluster algorithms can be found in \cite{Kawashima:1995a,Kawashima:1995b,Evertz:2000rk}.

\subsubsection{Cluster rules: Euclidean-time branch}

The first model under consideration is the Heisenberg model, whose Hamiltonian is given by
\begin{equation}
 \label{eq:heisenberg}
 H=J\sum_{\langle xy\rangle}\vec{S}_x\cdot\vec{S}_y \ .
\end{equation}
The model is anti-ferromagnetic for $J>0$ and ferromagnetic for $J<0$.
To construct the partition function, we split the Hamiltonian into commuting pieces and discretize Euclidean time by performing a Suzuki-Trotter decomposition.
Taking the $3$-direction as the quantization axis and denoting the eigenvalues of $S_{x}^3$ by $s_{x}\in\{\uparrow,\downarrow\}$, the non-vanishing plaquette weights $W[s_{x,k}s_{y,k},s_{x,k+1}s_{y,k+1}]$ on a bipartite lattice are calculated according to
\begin{subequations}
\label{eq:spinconfig_heisenberg}
\begin{align}
 W(1)&\equiv W[\uparrow\,\uparrow,\uparrow\,\uparrow]=W[\downarrow\,\downarrow,\downarrow\,\downarrow]=\exp\left(-\frac{\epsilon\beta J}{4}\right) \ , \\
 W(2)&\equiv W[\uparrow\,\downarrow,\uparrow\,\downarrow]=W[\downarrow\,\uparrow,\downarrow\,\uparrow]=\exp\left(\frac{\epsilon\beta J}{4}\right)\cosh\left(\frac{\epsilon\beta J}{2}\right) \ , \\
 W(3)&\equiv W[\uparrow\,\downarrow,\downarrow\,\uparrow]=W[\downarrow\,\uparrow,\uparrow\,\downarrow]=\exp\left(\frac{\epsilon\beta J}{4}\right)\sinh\left(\frac{\epsilon\beta|J|}{2}\right) \ ,
\end{align}
\end{subequations}
where $\epsilon=1/N_\tau$ determines the lattice spacing in the Euclidean time direction.
The loop-clusters are then generated by assigning bonds -- in the following denoted by $A$ or $A'$ (vertical parallel or anti-parallel), $B$ or $B'$ (horizontal parallel or anti-parallel) and $C$ or $C'$ (diagonal parallel or anti-parallel) -- with certain probabilities $p_{i,j}$ to any spin configuration.
Here, $i\in\{1,2,3,4\}$ corresponds to the spin configuration as introduced in \eqref{eq:spinconfig_heisenberg} whereas $j\in\{A,A',B,B',C,C'\}$ denotes the bond to be chosen.
For the anti-ferromagnetic Heisenberg model, the only non-vanishing probabilities are
\begin{subequations}
\label{eq:cluster_afm}
\begin{align}
 &p_{1,A}=p_{3,B'}=1 \ , \\
 &p_{2,A}=1-\tanh\left(\frac{\epsilon\beta J}{2}\right) \ , \\
 &p_{2,B'}=\tanh\left(\frac{\epsilon\beta J}{2}\right) \ .
\end{align}
\end{subequations}
We find that spins which are parallel in the vertical (Euclidean time) direction or anti-parallel in the horizontal (spatial) direction are bound together in loop-clusters.
On the other hand, the probabilities for the ferromagnetic Heisenberg model are given by
\begin{subequations}
\label{eq:cluster_fm}
\begin{align}
 &p_{1,A}=\exp\left(\frac{\epsilon\beta J}{2}\right)\cosh\left(\frac{\epsilon\beta J }{2}\right) \ , \\
 &p_{1,C}=\exp\left(\frac{\epsilon\beta J}{2}\right)\sinh\left(\frac{\epsilon\beta|J|}{2}\right) \ , \\ 
 &p_{2,A}=p_{3,C}=1 \ . 
\end{align}
\end{subequations}
In this case, only parallel spins are bound together either vertically or diagonally. 
The cluster rules for the Heisenberg model are summarized in Table~\ref{tab:cluster_euclid_heisenberg}.

\begin{table}[t]
\begin{tabular}{|C{2cm}|D{4.4cm}|D{4.4cm}|D{4.4cm}|}
 \hline
 plaquette & $\quad (1) \qquad \begin{bmatrix}\uparrow&\uparrow\\\uparrow&\uparrow\end{bmatrix}\  \text{or} \ \begin{bmatrix}\downarrow&\downarrow\\\downarrow&\downarrow\end{bmatrix}^{\phantom{X}}_{\phantom{X}}$ & $ \quad (2) \qquad \begin{bmatrix}\uparrow&\downarrow\\\uparrow&\downarrow\end{bmatrix}\  \text{or} \ \begin{bmatrix}\downarrow&\uparrow\\\downarrow&\uparrow\end{bmatrix}^{\phantom{X}}_{\phantom{X}}$ & $ \quad (3) \qquad \begin{bmatrix}\downarrow&\uparrow\\\uparrow&\downarrow\end{bmatrix}\  \text{or} \ \begin{bmatrix}\uparrow&\downarrow\\\downarrow&\uparrow\end{bmatrix}^{\phantom{X}}_{\phantom{X}}$  \\ \hline
 AFM bonds   & 
 $\ \begin{matrix*}[l] \phantom{0} \\ A \\ \phantom{0} \\ \phantom{0} \\ \phantom{0} \\ \phantom{0} \end{matrix*}$ \setlength{\unitlength}{1cm}
 \begin{picture}(0,0)\linethickness{0.4mm}\put(0.4,0.35){\line(0,1){0.6}}\put(1,0.35){\line(0,1){0.6}}\end{picture}
 $\qquad\qquad\qquad\quad\begin{matrix*}[c] \phantom{0} \\ 1 \\ \phantom{0} \\ \phantom{0} \\ \phantom{0} \\ \phantom{0} \end{matrix*}$ &  

 $\ \begin{matrix*}[l] \phantom{0} \\ A \\ \phantom{0} \\ \phantom{0} \\ B' \\ \phantom{0} \end{matrix*}$\setlength{\unitlength}{1cm}
 \begin{picture}(0,0)\linethickness{0.4mm}\put(0.4,0.35){\line(0,1){0.6}}\put(1,0.35){\line(0,1){0.6}}
                     \linethickness{0.4mm}\multiput(0.375,-0.15)(0.1,0){7}{\line(1,0){0.05}}\multiput(0.375,-0.75)(0.1,0){7}{\line(1,0){0.05}}\end{picture}
 $\qquad\qquad\begin{matrix*}[c] \phantom{0}\\ 1-\tanh\left(\tfrac{\epsilon\beta J}{2}\right) \\ \phantom{0} \\ \phantom{0} \\ \tanh\left(\tfrac{\epsilon\beta J}{2}\right) \\ \phantom{0} \end{matrix*}$ & 

 $\ \begin{matrix*}[l] \phantom{0} \\ \phantom{0} \\ \phantom{0} \\ \phantom{0} \\ B' \\ \phantom{0} \end{matrix*}$\setlength{\unitlength}{1cm}
 \begin{picture}(0,0)\linethickness{0.4mm}\multiput(0.375,-0.15)(0.1,0){7}{\line(1,0){0.05}}\multiput(0.375,-0.75)(0.1,0){7}{\line(1,0){0.05}}\end{picture}
 $\qquad\qquad\qquad\quad\begin{matrix*}[c] \phantom{0}\\ \phantom{0} \\ \phantom{0}\\ \phantom{0} \\ 1 \\ \phantom{0} \end{matrix*}$ \\ \hline

 FM bonds &  
 $\ \begin{matrix*}[l] \phantom{0} \\ A \\ \phantom{0} \\ \phantom{0} \\ C \\ \phantom{0} \end{matrix*}$ \setlength{\unitlength}{1cm}
 \begin{picture}(0,0)\linethickness{0.4mm}\put(0.4,0.35){\line(0,1){0.6}}\put(1,0.35){\line(0,1){0.6}}
 \thicklines\put(0.4,-0.75){\line(1,1){0.6}}\put(0.4,-0.15){\line(1,-1){0.6}}\end{picture}
 $\qquad\qquad\begin{matrix*}[c] \phantom{0} \\ e^{\frac{\epsilon\beta J}{2}}\cosh\left(\tfrac{\epsilon\beta|J|}{2}\right) \\ \phantom{0} \\ \phantom{0} \\ e^{\frac{\epsilon\beta J}{2}}\sinh\left(\tfrac{\epsilon\beta|J|}{2}\right) \\ \phantom{0} \end{matrix*}$ &  

 $\ \begin{matrix*}[l] \phantom{0} \\ A \\ \phantom{0} \\ \phantom{0} \\ \phantom{0} \\ \phantom{0} \end{matrix*}$ \setlength{\unitlength}{1cm}
 \begin{picture}(0,0)\linethickness{0.4mm}\put(0.4,0.35){\line(0,1){0.6}}\put(1,0.35){\line(0,1){0.6}}\end{picture}
 $\qquad\qquad\qquad\quad\begin{matrix*}[c] \phantom{0} \\ 1 \\ \phantom{0} \\ \phantom{0} \\ \phantom{0} \\ \phantom{0} \end{matrix*}$ &  

 $\ \begin{matrix*}[l] \phantom{0} \\ \phantom{0} \\ \phantom{0} \\ \phantom{0} \\ C \\ \phantom{0} \end{matrix*}$\setlength{\unitlength}{1cm}
 \begin{picture}(0,0)\thicklines\put(0.4,-0.75){\line(1,1){0.6}}\put(0.4,-0.15){\line(1,-1){0.6}}\end{picture}
 $\qquad\qquad\qquad\quad\begin{matrix*}[c] \phantom{0}\\ \phantom{0} \\ \phantom{0}\\ \phantom{0} \\ \ 1 \\ \phantom{0} \end{matrix*}$ \\ \hline
\end{tabular}
\caption{Euclidean-time cluster rules for the anti-ferromagnetic and ferromagnetic Heisenberg model. 
 Solid/dashed lines denote binding together parallel/anti-parallel spins, respectively, in the same loop-cluster.}
\label{tab:cluster_euclid_heisenberg}
\end{table}

\begin{table}[b]
\begin{tabular}{|C{2cm}|D{4.4cm}|D{3.5cm}|D{3.5cm}|D{3.5cm}|}
 \hline
 plaquette&$\quad (1) \quad \begin{bmatrix}\rightarrow&\rightarrow\\\rightarrow&\rightarrow\end{bmatrix}\  \text{or} \ \begin{bmatrix}\leftarrow&\leftarrow\\\leftarrow&\leftarrow\end{bmatrix}^{\phantom{X}}_{\phantom{X}}$ & $ \ (2) \ \begin{bmatrix}\rightarrow&\leftarrow\\\rightarrow&\leftarrow\end{bmatrix}\  \text{or} \ \begin{bmatrix}\leftarrow&\rightarrow\\\leftarrow&\rightarrow\end{bmatrix}^{\phantom{X}}_{\phantom{X}}$ & $ \ (3) \ \begin{bmatrix}\leftarrow&\rightarrow\\\rightarrow&\leftarrow\end{bmatrix}\  \text{or} \ \begin{bmatrix}\rightarrow&\leftarrow\\\leftarrow&\rightarrow\end{bmatrix}^{\phantom{X}}_{\phantom{X}}$& $\ (4) \ \begin{bmatrix}\leftarrow&\leftarrow\\\rightarrow&\rightarrow\end{bmatrix}\  \text{or} \ \begin{bmatrix}\rightarrow&\rightarrow\\\leftarrow&\leftarrow\end{bmatrix}^{\phantom{X}}_{\phantom{X}}$ \\ \hline
 XY bonds   & 
 $\ \begin{matrix*}[l] \phantom{0} \\ A \phantom{e^{\frac{k}{l}}}\\ \phantom{0} \\ \phantom{0} \\ B \phantom{e^{\frac{k}{l}}} \\ \phantom{0} \\ \phantom{0} \\ C \phantom{e^{\frac{k}{l}}} \\ \phantom{0} \end{matrix*}$ \setlength{\unitlength}{1cm}
 \begin{picture}(0,0)\linethickness{0.4mm}\put(0,0.95){\line(0,1){0.6}}\put(0.6,0.95){\line(0,1){0.6}}\put(0,0.3){\line(1,0){0.6}}\put(0,-0.2){\line(1,0){0.6}}\thicklines\put(0,-1.4){\line(1,1){0.6}}\put(0,-0.8){\line(1,-1){0.6}}\end{picture}
 $\qquad\quad\begin{matrix*}[c] \phantom{0} \\ e^{-\frac{\epsilon\beta J}{2}} \\ \phantom{0} \\ \phantom{0} \\ \tanh\left(\tfrac{\epsilon\beta J}{4}\right) \\ \phantom{0} \\ \phantom{0} \\ e^{-\frac{\epsilon\beta J}{2}}\tanh\left(\tfrac{\epsilon\beta J}{4}\right) \\ \phantom{0} \end{matrix*}$ &  

 $\ \begin{matrix*}[l] \phantom{0} \\ A \phantom{e^{\frac{k}{l}}} \\ \phantom{0} \\ \phantom{0} \\ \phantom{0} \\ \phantom{0} \\ \phantom{0} \\ \phantom{0} \\ \phantom{0} \end{matrix*}$ \setlength{\unitlength}{1cm}
 \begin{picture}(0,0)\linethickness{0.4mm}\put(0.4,0.95){\line(0,1){0.6}}\put(1,0.95){\line(0,1){0.6}}\end{picture}
 $\qquad\qquad\quad\begin{matrix*}[c] \phantom{0} \\ 1 \\ \phantom{0} \\ \phantom{0} \\ \phantom{e^{-\frac{k}{l}}} \\ \phantom{0} \\ \phantom{0} \\ \phantom{e^{-\frac{k}{l}}} \\ \phantom{0} \end{matrix*}$ &  

 $\ \begin{matrix*}[l] \phantom{0} \\ \phantom{e^{\frac{k}{l}}}\\ \phantom{0} \\ \phantom{0} \\ \phantom{e^{\frac{k}{l}}} \\ \phantom{0} \\ \phantom{0} \\ C \phantom{e^{\frac{k}{l}}} \\ \phantom{0} \end{matrix*}$ \setlength{\unitlength}{1cm}
 \begin{picture}(0,0)\thicklines\put(0.4,-1.4){\line(1,1){0.6}}\put(0.4,-0.8){\line(1,-1){0.6}}\end{picture}
 $\qquad\qquad\quad\begin{matrix*}[c] \phantom{0} \\ \phantom{e^{-\frac{k}{l}}} \\ \phantom{0} \\ \phantom{0} \\ \phantom{e^{-\frac{k}{l}}} \\ \phantom{0} \\ \phantom{0} \\ 1 \\ \phantom{0} \end{matrix*}$ &

 $\ \begin{matrix*}[l] \phantom{0} \\ \phantom{e^{\frac{k}{l}}}\\ \phantom{0} \\ \phantom{0} \\ B \phantom{e^{\frac{k}{l}}} \\ \phantom{0} \\ \phantom{0} \\ \phantom{e^{\frac{k}{l}}} \\ \phantom{0} \end{matrix*}$ \setlength{\unitlength}{1cm}
 \begin{picture}(0,0)\linethickness{0.4mm}\put(0.4,0.3){\line(1,0){0.6}}\put(0.4,-0.2){\line(1,0){0.6}}\end{picture}
 $\qquad\qquad\quad\begin{matrix*}[c] \phantom{0} \\ \phantom{e^{-\frac{k}{l}}} \\ \phantom{0} \\ \phantom{0} \\ 1\phantom{e^{-\frac{k}{l}}} \\ \phantom{0} \\ \phantom{0} \\ \phantom{e^{\frac{k}{l}}} \\ \phantom{0} \end{matrix*}$ \\ \hline
\end{tabular}
\caption{Euclidean-time cluster rules for the XY-model with the $1$--direction taken as the quantization axis.  Solid lines denote binding together parallel spins in the same loop-cluster.}
\label{tab:cluster_euclid_xy}
\end{table}

The second model Hamiltonian under consideration is given by
\begin{equation}
 \label{eq:xy}
 H=-J\sum_{\langle xy\rangle}\left(S^1_xS^{1}_y+S^2_xS^{2}_y\right)=-\frac{J}{2}\sum_{\langle xy\rangle}\left(S^+_xS^{-}_y+S^-_xS^{+}_y\right) \ ,
\end{equation}
corresponding to the (ferromagnetic) quantum XY-model for $J>0$.
Performing the same steps as for the Heisenberg model but taking the $1$-direction as the quantization axis and denoting the eigenvalues of $S_{x}^1$ by $s_{x}\in\{\rightarrow,\leftarrow\}$, on a bipartite lattice we obtain the following non-vanishing plaquette weights
\begin{subequations}
\label{eq:spinconfig_xy}
\begin{align}
 W(1)&\equiv W[\rightarrow\,\rightarrow,\rightarrow\,\rightarrow]=W[\leftarrow\,\leftarrow,\leftarrow\,\leftarrow]=\exp\left(\frac{\epsilon\beta J}{4}\right)\cosh\left(\frac{\epsilon\beta J}{4}\right) \ , \\
 W(2)&\equiv W[\rightarrow\,\leftarrow,\rightarrow\,\leftarrow]=W[\leftarrow\,\rightarrow,\leftarrow\,\rightarrow]=\exp\left(-\frac{\epsilon\beta J}{4}\right)\cosh\left(\frac{\epsilon\beta J}{4}\right) \ , \\
 W(3)&\equiv W[\rightarrow\,\leftarrow,\leftarrow\,\rightarrow]=W[\leftarrow\,\rightarrow,\rightarrow\,\leftarrow]=\exp\left(-\frac{\epsilon\beta J}{4}\right)\sinh\left(\frac{\epsilon\beta J }{4}\right) \ , \\
 W(4)&\equiv W[\rightarrow\,\rightarrow,\leftarrow\,\leftarrow]=W[\leftarrow\,\leftarrow,\rightarrow\,\rightarrow]=\exp\left(\frac{\epsilon\beta J}{4}\right)\sinh\left(\frac{\epsilon\beta J }{4}\right) \ .
\end{align}
\end{subequations}
The corresponding cluster rules, which bind together only parallel spins and which are summarized in Table~\ref{tab:cluster_euclid_xy}, are then determined by
\begin{subequations}
\label{eq:cluster_xy}
\begin{align}
 &p_{1,A}=\exp\left(-\frac{\epsilon\beta J}{2}\right) \ , \\
 &p_{1,B}=\tanh\left(\frac{\epsilon\beta J}{4}\right) \ , \\
 &p_{1,C}=\exp\left(-\frac{\epsilon\beta J}{2}\right)\tanh\left(\frac{\epsilon\beta J}{4}\right) \ , \\ 
 &p_{2,A}=p_{3,C}=p_{4,B}=1 \ . 
\end{align}
\end{subequations}

\subsubsection{Cluster rules: Real-time branch}

To derive the cluster rules for the real-time branch of the Konstantinov-Perel' contour, we have to consider \eqref{eq:Lindblad_pairs} for the various measurement processes.
More specifically, we have to further investigate the averaged measurement results \eqref{eq:cluster_total}, \eqref{eq:cluster_x} and \eqref{eq:cluster_rl}.
For all these measurement processes, it follows from \eqref{eq:Lindblad_pairs} that the spin configurations and loop-clusters that contribute to the path integral are identical on both branches of the real-time path, $s_{x,k}=s'_{x,k}$.
Accordingly, we can restrict our attention to deriving the cluster rules on the forward branch of the contour, which are then summarized in Table~\ref{tab:cluster_real}.

\begin{table}[t]
\begin{tabular}{|C{2cm}|D{3.5cm}|D{3.5cm}|D{3.5cm}|D{3.5cm}|}
 \hline
 plaquette&$\quad (1) \quad \begin{bmatrix}\uparrow&\uparrow\\\uparrow&\uparrow\end{bmatrix}\  \text{or} \ \begin{bmatrix}\downarrow&\downarrow\\\downarrow&\downarrow\end{bmatrix}^{\phantom{X}}_{\phantom{X}}$ & $ \quad (2) \quad \begin{bmatrix}\uparrow&\downarrow\\\uparrow&\downarrow\end{bmatrix}\  \text{or} \ \begin{bmatrix}\downarrow&\uparrow\\\downarrow&\uparrow\end{bmatrix}^{\phantom{X}}_{\phantom{X}}$ & $ \quad (3) \quad \begin{bmatrix}\downarrow&\uparrow\\\uparrow&\downarrow\end{bmatrix}\  \text{or} \ \begin{bmatrix}\uparrow&\downarrow\\\downarrow&\uparrow\end{bmatrix}^{\phantom{X}}_{\phantom{X}}$&$\quad (4) \quad \begin{bmatrix}\downarrow&\downarrow\\\uparrow&\uparrow\end{bmatrix}\  \text{or} \ \begin{bmatrix}\uparrow&\uparrow\\\downarrow&\downarrow\end{bmatrix}^{\phantom{X}}_{\phantom{X}}$   \\ \hline
 $O^{(1)}$ bonds &
 $\ \begin{matrix*}[l] \phantom{0} \\ A \\ \phantom{0} \\ \phantom{0} \\ C \\ \phantom{C'} \end{matrix*}$ \setlength{\unitlength}{1cm}
 \begin{picture}(0,0)\linethickness{0.4mm}\put(0.4,0.35){\line(0,1){0.6}}\put(1,0.35){\line(0,1){0.6}}
 \thicklines\put(0.4,-0.75){\line(1,1){0.6}}\put(0.4,-0.15){\line(1,-1){0.6}}\end{picture}
 $\qquad\qquad\ \ \begin{matrix*}[c] \phantom{0} \\ 1-\tfrac{\epsilon\gamma}{2} \\ \phantom{0} \\ \phantom{0} \\ \tfrac{\epsilon\gamma}{2} \\ \phantom{0} \end{matrix*}$ &  
 $\ \begin{matrix*}[l] \phantom{0} \\ A \\ \phantom{0} \\ \phantom{0} \\ \phantom{C} \\ \phantom{C'} \end{matrix*}$ \setlength{\unitlength}{1cm}
 \begin{picture}(0,0)\linethickness{0.4mm}\put(0.4,0.35){\line(0,1){0.6}}\put(1,0.35){\line(0,1){0.6}}\end{picture}
 $\qquad\qquad\ \begin{matrix*}[c] \phantom{0} \\ 1 \\ \phantom{0} \\ \phantom{0} \\ \phantom{1-\tfrac{\epsilon\gamma}{2-\epsilon\gamma}} \\ \phantom{0} \end{matrix*}$ &  
 $\ \begin{matrix*}[l] \phantom{0} \\ \phantom{A} \\ \phantom{0} \\ \phantom{0} \\ C \\ \phantom{0} \end{matrix*}$\setlength{\unitlength}{1cm}
 \begin{picture}(0,0)\thicklines\put(0.4,-0.75){\line(1,1){0.6}}\put(0.4,-0.15){\line(1,-1){0.6}}\end{picture}
 $\qquad\qquad\quad\begin{matrix*}[c] \phantom{0}\\ \phantom{1-\tfrac{\epsilon\gamma}{2}} \\ \phantom{0}\\ \phantom{0} \\ \ 1 \\ \phantom{0} \end{matrix*}$ &
 \begin{center}not possible\end{center}\\\hline

 $O^{(2)}$ bonds &
 $\ \begin{matrix*}[l] \phantom{0} \\ A \\ \phantom{0} \\ \phantom{0} \\ C \\ \phantom{0} \\ \phantom{0} \\ \phantom{C'} \\ \phantom{0} \end{matrix*}$ \setlength{\unitlength}{1cm}
 \begin{picture}(0,0)\linethickness{0.4mm}\put(0.4,0.95){\line(0,1){0.6}}\put(1,0.95){\line(0,1){0.6}}
 \thicklines\put(0.4,-0.2){\line(1,1){0.6}}\put(0.4,0.4){\line(1,-1){0.6}}\end{picture}
 $\qquad\qquad \begin{matrix*}[c] \phantom{0} \\ 1-\tfrac{\epsilon\gamma}{2-\epsilon\gamma} \\ \phantom{0} \\ \phantom{0} \\ \tfrac{\epsilon\gamma}{2-\epsilon\gamma} \\ \phantom{0} \\ \phantom{0} \\ \phantom{0} \\ \phantom{0} \end{matrix*}$ & 
 $\ \begin{matrix*}[l] \phantom{0} \\ A \\ \phantom{0} \\ \phantom{0} \\ \phantom{\tfrac{X}{Y}} \\ \phantom{0} \\ \phantom{0} \\ C' \\ \phantom{0} \end{matrix*}$ \setlength{\unitlength}{1cm}
 \begin{picture}(0,0)\linethickness{0.4mm}\put(0.4,0.95){\line(0,1){0.6}}\put(1,0.95){\line(0,1){0.6}}
 \thicklines\multiput(0.4,-1.25)(0.05,0.05){13}{\line(1,0){0.025}}\multiput(0.4,-0.65)(0.05,-0.05){13}{\line(1,0){0.025}}\end{picture}
 $\qquad\qquad\begin{matrix*}[c] \phantom{0} \\ 1-\tfrac{\epsilon\gamma}{2-\epsilon\gamma} \\ \phantom{0} \\ \phantom{0} \\ \phantom{0} \\ \phantom{0} \\ \phantom{0} \\ \tfrac{\epsilon\gamma}{2-\epsilon\gamma} \\ \phantom{0} \end{matrix*}$ & 
 $\ \begin{matrix*}[l] \phantom{0} \\ \phantom{A} \\ \phantom{0} \\ \phantom{0} \\ C \\ \phantom{0} \\ \phantom{0} \\ \phantom{0} \\ \phantom{0} \end{matrix*}$ \setlength{\unitlength}{1cm}
 \begin{picture}(0,0)\thicklines\put(0.4,-0.2){\line(1,1){0.6}}\put(0.4,0.4){\line(1,-1){0.6}}\end{picture}
 $\qquad\qquad\qquad\begin{matrix*}[c] \phantom{0} \\ \phantom{0} \\ \phantom{0} \\ \phantom{0} \\ 1 \\ \phantom{0} \\ \phantom{0} \\ \phantom{0} \\ \phantom{0} \end{matrix*}$ & 
 $\ \begin{matrix*}[l] \phantom{0} \\ \phantom{A} \\ \phantom{\tfrac{X}{Y}} \\ \phantom{0} \\ \phantom{C} \\ \phantom{0} \\ \phantom{0} \\ C' \\ \phantom{0} \end{matrix*}$ \setlength{\unitlength}{1cm}
 \begin{picture}(0,0)\thicklines\multiput(0.4,-1.25)(0.05,0.05){13}{\line(1,0){0.025}}\multiput(0.4,-0.65)(0.05,-0.05){13}{\line(1,0){0.025}}\end{picture}
 $\qquad\qquad\begin{matrix*}[c] \phantom{0} \\ \phantom{1-\tfrac{\epsilon\gamma}{2-\epsilon\gamma}} \\ \phantom{0} \\ \phantom{0} \\ \phantom{0} \\ \phantom{0} \\ \phantom{0} \\ 1 \\ \phantom{0} \end{matrix*}$ \\\hline

 $O^{(3)}$ bonds & 
 $\ \begin{matrix*}[l] \phantom{0} \\ A \\ \phantom{0} \\ \phantom{0} \\ \phantom{C'} \\ \phantom{0} \end{matrix*}$ \setlength{\unitlength}{1cm}
 \begin{picture}(0,0)\linethickness{0.4mm}\put(0.4,0.35){\line(0,1){0.6}}\put(1,0.35){\line(0,1){0.6}}\end{picture}
 $\qquad\qquad \begin{matrix*}[c] \phantom{0} \\ 1 \\ \phantom{0} \\ \phantom{0} \\ \phantom{1-\tfrac{\epsilon\gamma}{2-\epsilon\gamma}} \\ \phantom{0} \end{matrix*}$ &  
 $\ \begin{matrix*}[l] \phantom{0} \\ A \\ \phantom{0} \\ \phantom{0} \\ C' \\ \phantom{0} \end{matrix*}$ \setlength{\unitlength}{1cm}
 \begin{picture}(0,0)\linethickness{0.4mm}\put(0.4,0.35){\line(0,1){0.6}}\put(1,0.35){\line(0,1){0.6}}
 \thicklines\multiput(0.4,-0.75)(0.05,0.05){13}{\line(1,0){0.025}}\multiput(0.4,-0.15)(0.05,-0.05){13}{\line(1,0){0.025}}\end{picture}
 $\qquad\qquad \ \begin{matrix*}[c] \phantom{0} \\ 1-\tfrac{\epsilon\gamma}{2} \\ \phantom{0} \\ \phantom{0} \\ \tfrac{\epsilon\gamma}{2} \\ \phantom{0} \end{matrix*}$ &  
 \begin{center}not possible\end{center}&
 $\ \begin{matrix*}[l] \phantom{0} \\ \phantom{0} \\ \phantom{0} \\ \phantom{0} \\ C' \\ \phantom{0} \end{matrix*}$\setlength{\unitlength}{1cm}
 \begin{picture}(0,0)\thicklines\multiput(0.4,-0.75)(0.05,0.05){13}{\line(1,0){0.025}}\multiput(0.4,-0.15)(0.05,-0.05){13}{\line(1,0){0.025}}
\end{picture}
 $\qquad\qquad\qquad\begin{matrix*}[c] \phantom{0}\\ \phantom{0} \\ \phantom{0}\\ \phantom{0} \\ \ 1 \\ \phantom{0} \end{matrix*}$  \\ \hline
\end{tabular}
\caption{Real-time cluster rules for the different measurement processes $O^{(i)}$ with $i\in\{1,2,3\}$, with the $3$--direction taken as the quantization axis.  
Solid/dashed lines denote binding together parallel/anti-parallel spins, respectively, in the same loop-cluster.}
\label{tab:cluster_real}
\end{table}

For the measurement process corresponding to the total spin $O^{(1)}=\vec{S}^2$ of two neighboring quantum spins, we find 
\begin{align}
 &\bracket{s_{x,k-1}s_{y,k-1}s'_{x,k-1}s'_{y,k-1}}{(1-\epsilon\gamma)\mathbb{1}\otimes\mathbb{1}+\epsilon\gamma\widetilde{P}_k}{s_{x,k}s_{y,k}s'_{x,k}s'_{y,k}}= \nonumber \\
 &\bracket{s_{x,k-1}s_{y,k-1}}{(1-\epsilon\gamma)\mathbb{1}+\epsilon\gamma P_1}{s_{x,k}s_{y,k}}=\left(1-\frac{\epsilon\gamma}{2}\right)\delta_{s_{x,k-1},s_{x,k}}\delta_{s_{y,k-1},s_{y,k}}+\frac{\epsilon\gamma}{2}\delta_{s_{x,k-1},s_{y,k}}\delta_{s_{y,k-1},s_{x,k}} \ .
\end{align}
Employing the same notation as before, this corresponds to cluster rules for binding together parallel spins both vertically (real time) and diagonally
\begin{subequations}
\label{eq:cr_total}
\begin{align}
 &p_{1,A}=1-\frac{\epsilon\gamma}{2} \ , \\
 &p_{1,C}=\frac{\epsilon\gamma}{2} \ , \\ 
 &p_{2,A}=p_{3,C}=1 \ .
\end{align}
\end{subequations}
For the measurement process $O^{(2)}=S_x^1S_y^1$, on the other hand, we obtain
\begin{align}
 &\bracket{s_{x,k-1}s_{y,k-1}s'_{x,k-1}s'_{y,k-1}}{(1-\epsilon\gamma)\mathbb{1}\otimes\mathbb{1}+\epsilon\gamma\widetilde{P}_k}{s_{x,k}s_{y,k}s'_{x,k}s'_{y,k}}= \nonumber \\
 &\bracket{s_{x,k-1}s_{y,k-1}}{(1-\epsilon\gamma)\mathbb{1}+\epsilon\gamma P_\parallel}{s_{x,k}s_{y,k}}=\left(1-\frac{\epsilon\gamma}{2}\right)\delta_{s_{x,k-1},s_{x,k}}\delta_{s_{y,k-1},s_{y,k}}+\frac{\epsilon\gamma}{2}\delta_{s_{x,k-1},-s_{x,k}}\delta_{s_{y,k-1},-s_{y,k}} \ .
\end{align}
The cluster rules for this process are more intricate.
Most notably, this measurement process allows for plaquette configurations of the form $W(4)=W[\uparrow\,\uparrow,\downarrow\,\downarrow]=W[\downarrow\,\downarrow,\uparrow\,\uparrow]$, which had a vanishing plaquette weight in the Euclidean time branch of the Heisenberg model. 
Accordingly, the cluster rules which bind together parallel spins both vertically and diagonally whereas anti-parallel spins are bound together in the diagonal direction are determined by
\begin{subequations}
\label{eq:cr_x}
\begin{align}
 &p_{1,A}=p_{2,A}=1-\frac{\epsilon\gamma}{2-\epsilon\gamma} \ , \\
 &p_{1,C}=p_{2,C'}=\frac{\epsilon\gamma}{2-\epsilon\gamma} \ , \\
 &p_{3,C}=p_{4,C'}=1 \ .
\end{align}
\end{subequations}
Finally, the measurement process $O^{(3)}=S_x^+S_y^+ + S_x^-S_y^-$ results in
\begin{align}
 &\bracket{s_{x,k-1}s_{y,k-1}s'_{x,k-1}s'_{y,k-1}}{(1-\epsilon\gamma)\mathbb{1}\otimes\mathbb{1}+\epsilon\gamma\widetilde{P}_k}{s_{x,k}s_{y,k}s'_{x,k}s'_{y,k}}= \nonumber \\
 &\bracket{s_{x,k-1}s_{y,k-1}}{(1-\epsilon\gamma)\mathbb{1}+\epsilon\gamma (P_++P_0)}{s_{x,k}s_{y,k}}=\left(1-\frac{\epsilon\gamma}{2}\right)\delta_{s_{x,k-1},s_{x,k}}\delta_{s_{y,k-1},s_{y,k}}+\frac{\epsilon\gamma}{2}\delta_{s_{x,k-1},-s_{y,k}}\delta_{s_{y,k-1},-s_{x,k}} \ .
\end{align}
In fact, we find a vanishing plaquette weight $W(3)=W[\uparrow\,\downarrow,\downarrow\,\uparrow]=W[\downarrow\,\uparrow,\uparrow\,\downarrow]=0$ on the real-time branch for this process.
This is also reflected in the cluster rules, which bind together parallel spins vertically whereas anti-parallel spins are bound together diagonally
\begin{subequations}
\label{eq:cr_rl}
\begin{align}
 &p_{1,A}=p_{4,C'}=1 \ , \\
 &p_{2,A}=1-\frac{\epsilon\gamma}{2} \ , \\
 &p_{2,C'}=\frac{\epsilon\gamma}{2} \ . 
\end{align}
\end{subequations}

Concluding, we remark that these cluster rules are dedicated to measuring $O^{(i)}$ with $i\in\{1,2,3\}$ of spin systems which have been quantized in the $3$-direction. 
We consider, however, the XY-model for spins which are quantized in the $1$-direction.
Accordingly, these cluster rules then correspond to different measurement processes:
In fact, the cluster rules \eqref{eq:cr_total} still correspond to measuring the total spin $\vec{S}^2$.
On the other hand, the cluster rules \eqref{eq:cr_x} are then related to measuring $S_x^{2}S_y^{2}$ or $S_x^{3}S_y^{3}$,
and the cluster rules \eqref{eq:cr_rl} correspond to the measurement process $S_x^{2}S_y^{2}-S_x^{3}S_y^{3}$.

\section{Heisenberg ferromagnet initial state}
\label{app:results_fm}

After discussing the measurement-driven real-time dynamics in the anti-ferromagnetic Heisenberg model in great detail in Sec.~\ref{sec:results_afm}, we now consider an initial density matrix $\rho_0$ corresponding to the ferromagnetic Heisenberg model \eqref{eq:heisenberg}, where we choose again the $3$-direction as the quantization axis.
The ensemble of initial states is then prepared by means of the Euclidean-time cluster rules \eqref{eq:cluster_fm}.
We then again study the time-dependence of the Fourier modes \eqref{eq:fourier_modes} for the different measurement processes $O^{(i)}$ with $i\in\{1,2,3\}$.
As the measurement processes are independent of the chosen initial state, we again find the conservation of the $(0,0)$-mode -- which now corresponds to the order parameter of the system -- for $O^{(1)}$ and the conservation of the $(\pi,\pi)$-mode for $O^{(3)}$.
On the other hand, the measurement process $O^{(2)}$ does again not conserve any of the Fourier modes.

Applying the same methods as before and assuming that the system is prepared at low temperature $T\to0$ at the initial time $t_0$, we can again calculate the final equilibrium values analytically.
The ground state has total spin $\mathcal{M}=N^2/2$ and is $(2\mathcal{M}+1)$--fold degenerate, so that we find
\begin{equation}
 \label{eq:initial_fm}
 A^{(1)}(p=(0,0))=\langle|S(p=(0,0))|^2\rangle(t_0)=\frac{\mathcal{M}(\mathcal{M}+1)}{3}=\frac{N^2(N^2+2)}{12} 
\end{equation}
for the measurement process $O^{(1)}$.
Due to the fact that the $(0,0)$-mode is exactly conserved during the time evolution, the system is driven to a final equilibrium ensemble for which the $3$-component of the uniform magnetization is uniformly distributed among the $2\mathcal{M}+1$ sectors with $M\in\{-\mathcal{M},...,\mathcal{M}\}$.
In fact, the corresponding spin-spin-correlation function in any sector of $M$ is given by
\begin{equation}
 \langle S_{x}^3S_{y}^3\rangle_{M}=\frac{[N^4-(2M)^2]\delta_{x,y}-[N^2-(2M)^2]}{4N^2(N^2-1)} \ .
\end{equation}
Evaluating \eqref{eq:fourier_modes} and averaging over the $2\mathcal{M}+1$ sectors of uniform magnetization then gives
\begin{subequations}
\label{eq:final_total_fm}
\begin{align}
 A^{(1)}(p=(0,0))&=\frac{N^2(N^2+2)}{12} \ , \\
 A^{(1)}(p\neq(0,0))&=\frac{N^2}{6} \ .
\end{align}
\end{subequations}
As these values also correspond to their equilibrium values at the initial time $t_0$, we do not observe any interesting dynamics upon applying measurement process $O^{(1)}$ to the ferromagnetic Heisenberg model at $T\to0$.

\begin{figure}[t]
 \includegraphics[width=\columnwidth]{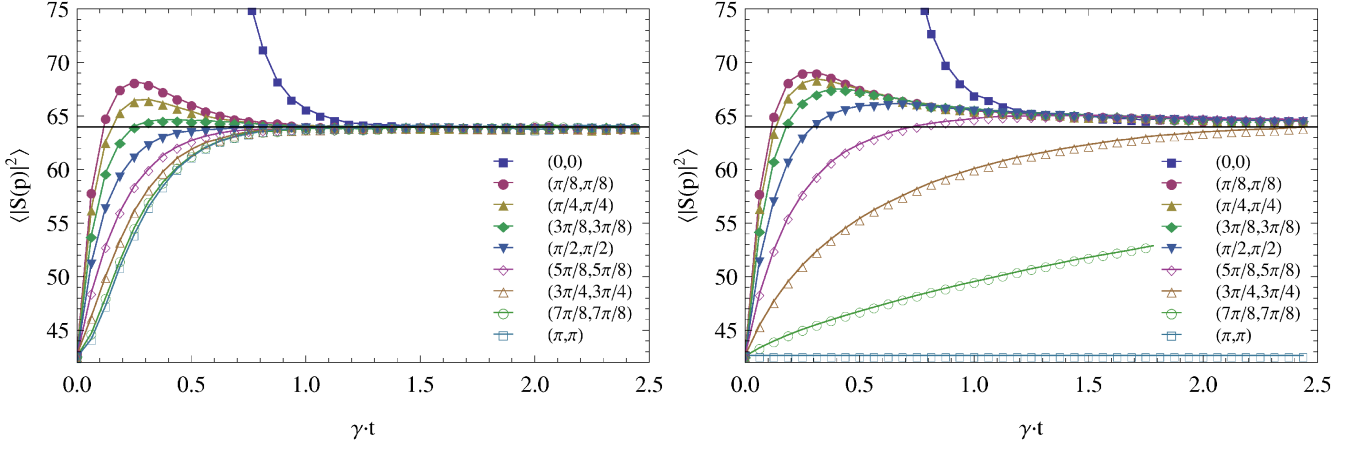}
 \caption{\label{fig:evolution_fm}
 [Color online] (Heisenberg ferromagnet initial state) 
 Time evolution of certain Fourier modes $\langle|S(p)|^2\rangle(t)$ for a short time interval for the measurement processes $O^{(2)}$ ({\it left}) and $O^{(3)}$ ({\it right}).
 The error bars are again of the order of the symbol sizes and the lines are included to guide the eye.
 We initialize a ferromagnetic Heisenberg model at low temperatures such that the $(0,0)$-mode is very large whereas the $(\pi,\pi)$-mode is small.
 The remaining parameters are $4N_\tau=512$, $L=16a$, $\beta J = 5L/2a=40$ and $\epsilon\gamma=0.05$.
 In order to determine the time evolution, we performed $10^{6}$ Monte Carlo measurements.
 The horizontal lines corresponds to the analytically derived final equilibrium values \eqref{eq:final_x_fm} and \eqref{eq:final_rl_fm}, respectively.}
\end{figure}

For the measurement process $O^{(2)}$, on the other hand, the final equilibrium corresponds again to a state where each of the $2^{N^2}$ spin states is equally probable.
Accordingly, all Fourier modes again take the same final value
\begin{equation}
 \label{eq:final_x_fm}
 A^{(2)}(p)=\frac{N^2}{4} \ .
\end{equation}
We find that all Fourier modes show a rapid equilibration towards these new equilibrium values, as shown in Fig.~\ref{fig:evolution_fm}. 
Note again the overshooting of several Fourier modes which can be traced back to \eqref{eq:fourier_sum} but the absence of a non-trivial attractor.

Finally, for the measurement process $O^{(3)}$ we cannot use the same reasoning as in the anti-ferromagnet as the probability $p(M_s)$ of being in a certain $M_s$-sector at $t_0$ is unknown.
The numerical results, however, indicate that the conserved $(\pi,\pi)$-mode takes the value
\begin{equation}
 A^{(3)}(p=(\pi,\pi))=\langle|S(p=(\pi,\pi))|^2\rangle(t_0)\simeq\frac{N^2}{6} \ .
\end{equation}
Moreover, as it is still true that all Fourier modes $A^{(3)}(p\neq(\pi,\pi))$ reach the same final equilibrium value, we conclude
\begin{subequations}
\begin{align}
 A^{(3)}(p=(\pi,\pi))&=\langle|S(p=(\pi,\pi))|^2\rangle(t_0)\simeq\frac{N^2}{6} \ , \\
 \label{eq:final_rl_fm}
 A^{(3)}(p\neq(\pi,\pi))&=\frac{1}{N^2-1}\left[\frac{N^4}{4}-\langle|S(p=(\pi,\pi))|^2\rangle(t_0)\right]\simeq\frac{N^2(3N^2-2)}{12(N^2-1)} \ .
\end{align}
\end{subequations}
Due to the appearance of slow Fourier modes in the vicinity of the $(\pi,\pi)$-mode, we again find a non-trivial attractor for the dynamics which is driven by measurement process $O^{(3)}$, as shown in Fig.~\ref{fig:evolution_fm}.
As a consequence, we can again simultaneously fit the Fourier modes in the vicinity of the $(\pi,\pi)$-mode for measurement process $O^{(3)}$ upon taking into account the analytically predicted attractor behavior.
The attraction rates, at which the different Fourier modes are driven towards the attractor $\mathcal{A}(t)$, are then again found to depend quadratically on the momentum value \eqref{eq:mom_dep_o3} with $c_{3}=1.22(3)$ and $r_{3}=2.04(5)$.

Finally, we studied again the time-dependence of $\langle|S(p=(0,0))|^2\rangle/L^2$, whose large value \eqref{eq:initial_fm} $\sim\mathcal{O}(L^2)$ indicates spontaneous symmetry breaking of the $SU(2)$ spin symmetry at zero temperature.
On the other hand, after driving the system with the measurement processes $O^{(2)}$ and $O^{(3)}$ to its new equilibrium state, we find that it becomes volume-independent suggesting the restoration of the $SU(2)$ spin symmetry.
As a matter of fact, the Binder ratio $B_4(t)$ shows a very similar behavior as for the anti-ferromagnetic initial state, which indicates that the phase transition is again completed only after an infinite amount of time.

\section{XY-model initial state}
\label{app:results_xy}

In this section, we briefly discuss the dynamics for an initial density matrix $\rho_0$ corresponding to the ferromagnetic quantum XY-model \eqref{eq:xy}.
We note that the anti-ferromagnetic and ferromagnetic XY-model are unitarily equivalent on a bipartite lattice in the absence of a magnetic field.
In order to have direct access to the order parameter of the system, we now choose the $1$-direction as the quantization axis.
The ensemble of initial states is then prepared by means of the Euclidean-time cluster rules \eqref{eq:cluster_xy}.
Due to the different quantization axis, the Fourier modes are now defined according to
\begin{equation}
 S(p)=\sum_{x}{\exp\left(ipx\right)S_x^1}=\sum_{x}{\exp\left(ip_1x_1+ip_2x_2\right)S_x^1} \ .
\end{equation}
We also note that the measurement processes have a slightly different meaning than before as already discussed previously.
Nevertheless, we have again the conservation of the $(0,0)$-mode for $O^{(1)}$ and the conservation of the $(\pi,\pi)$-mode for $O^{(3)}$.
On the other hand, the measurement process $O^{(2)}$ does again not conserve any of the Fourier modes.

Preparing an initial state at low temperature $T\to0$, we study the time-dependence of the Fourier modes for the different measurement processes.
For this initial state, finite-volume chiral perturbation theory predicts
\begin{equation}
 \label{eq:analytic_order_xy}
 \langle|S(p=(0,0)|^2\rangle(t_0)\simeq\frac{\mathcal{M}^2 L^4}{2}\sum_{n} a_n\left(\frac{c}{\rho L}\right)^n  \ ,
\end{equation}
where the first three coefficients are given by $a_0=1$, $a_1=0.45157$ and $a_2=0.030793$ \cite{Gockeler:1990ik,Gockeler:1990zn,Hasenfratz:1993}.
The low-energy parameters are the magnetization density $\mathcal{M}=0.43561(1)/a^2$, the spin stiffness $\rho=0.26974(5)J$ and the spin-wave velocity $c=1.1347(2)Ja$ \cite{Jiang:2011,Gerber:2011ya}.

Again, the final equilibrium for the measurement process $O^{(2)}$ corresponds to a state where each of the $2^{N^2}$ spin states are equally probable so that all Fourier modes take the same value
\begin{equation}
 \label{eq:final_x_xy}
 A^{(2)}(p)=\frac{N^2}{4} \ .
\end{equation}
As in the Heisenberg model, we then find that all Fourier modes show a rapid equilibration towards these new equilibrium values without generating a non-trivial attractor.

\begin{figure}[t]
 \includegraphics[width=\columnwidth]{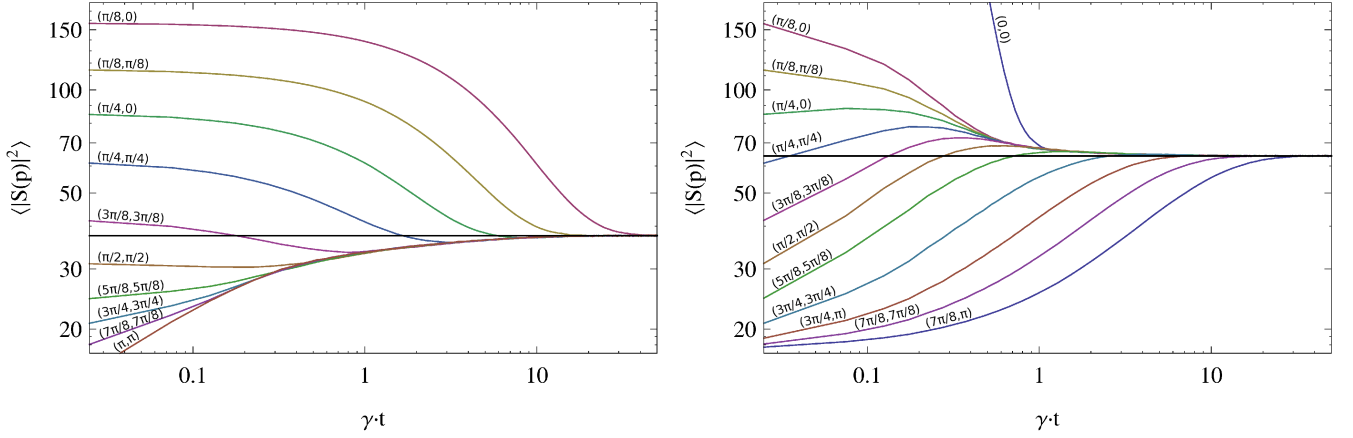}
 \caption{\label{fig:evolution_xy} 
 [Color online] (XY-model initial state) 
 Time evolution of certain Fourier modes $\langle|S(p)|^2\rangle(t)$ for a long time interval for the measurement processes $O^{(1)}$ ({\it left}) and $O^{(3)}$ ({\it right}).
 The error bars are of the order of the line width.
 We initialize the XY-model at low temperatures such that the $(0,0)$-mode is large whereas the $(\pi,\pi)$-mode is small.
 The remaining parameters are $4N_\tau=512$, $L=16a$, $\beta J = 5L/2a=40$ and $\epsilon\gamma=0.05$.
 In order to determine the time evolution, we performed $10^{6}$ Monte Carlo measurements.
 The horizontal lines corresponds to the final equilibrium values \eqref{eq:final_tot_xy} and \eqref{eq:final_rl_xy}, respectively.}
\end{figure}

For the measurement processes $O^{(1)}$ and $O^{(3)}$, however, we are not able to predict the final equilibrium values $A(p)$ analytically.
Nevertheless, it is still true that all Fourier modes except the conserved one converge towards the same final equilibrium value.
Accordingly, we have for the measurement process $O^{(1)}$
\begin{subequations}
\begin{align}
 A^{(1)}(p=(0,0))&=\langle|S(p=(0,0))|^2\rangle(t_0)\ , \\
 \label{eq:final_tot_xy}
 A^{(1)}(p\neq(0,0))&=\frac{1}{N^2-1}\left[\frac{N^4}{4}-\langle|S(p=(0,0))|^2\rangle(t_0)\right] \ ,
\end{align}
\end{subequations}
with $\langle|S(p=(0,0))|^2\rangle(t_0)$ being given by \eqref{eq:analytic_order_xy}.
On the other hand, for the measurement process $O^{(3)}$ we find
\begin{subequations}
\begin{align}
 A^{(3)}(p=(\pi,\pi))&=\langle|S(p=(\pi,\pi))|^2\rangle(t_0)\ , \\
 \label{eq:final_rl_xy}
 A^{(3)}(p\neq(\pi,\pi))&=\frac{1}{N^2-1}\left[\frac{N^4}{4}-\langle|S(p=(\pi,\pi))|^2\rangle(t_0)\right] \ ,
\end{align}
\end{subequations}
where $\langle|S(p=(\pi,\pi))|^2\rangle(t_0)$ needs to be determined numerically.
In both cases, we again observe slow Fourier modes in the vicinity of the conserved Fourier mode, as shown in Fig.~\ref{fig:evolution_xy}. 
As a consequence, a non-trivial attractor $\mathcal{A}(t)$ is formed towards which all equilibrating Fourier modes except the slowest one are driven before the final equilibrium value $A(p)$ is reached.
The attraction rates, at which the different Fourier modes are driven towards the attractor, are then again found to depend quadratically on the momentum value \eqref{eq:mom_dep_o1} with $c_{1}=1.13(2)$ and $r_{1}=2.04(4)$ as well as \eqref{eq:mom_dep_o3} with $c_{3}=1.19(3)$ and $r_{3}=2.02(5)$.

\begin{figure}[t]
 \includegraphics[width=0.95\columnwidth]{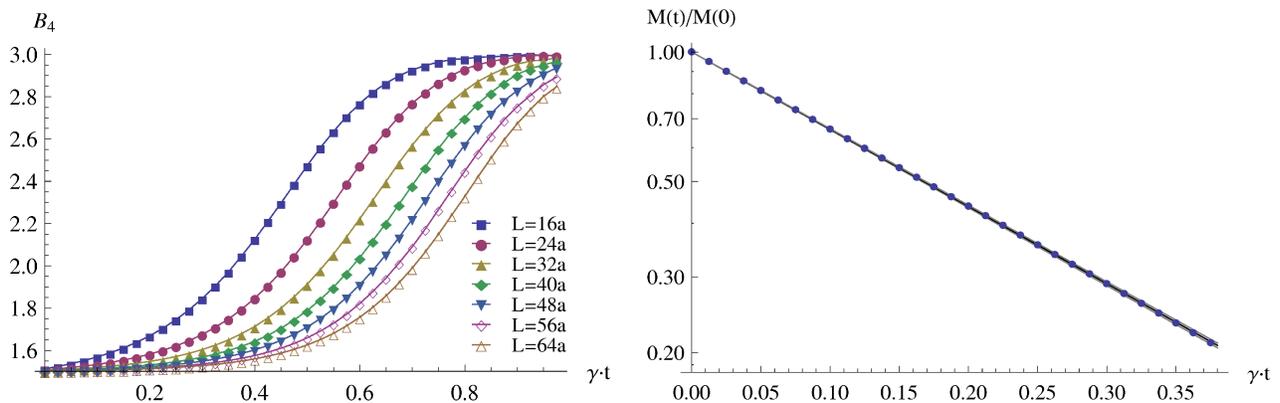}
 \caption{\label{fig:phase_transition_xy} 
 [Color online] (XY-model initial state) 
 {\it Left:}  Time-dependent Binder ratio $B_4(t)$ for the measurement process $O^{(3)}$ for different values of $L/a$ along with variable $\beta J=3L/4a$. 
 The lines are included to guide the eye.
 {\it Right:} Time-dependent magnetization density $\mathcal{M}(t)/\mathcal{M}(0)$ for the measurement process $O^{(3)}$ on a logarithmic plot along with an exponential fit of the data.
 In both cases, the error bars are of the order of the symbol sizes.
 We emphasize that the results for the measurement process $O^{(2)}$ are essentially identical.}
\end{figure}

Finally, we investigate the time-dependence of $\langle|S(p=(0,0))|^2\rangle/L^2$, whose large value \eqref{eq:analytic_order_xy} $\sim\mathcal{O}(L^2)$ is a consequence of the quasi long-range order in the two-dimensional XY-model below the Kosterlitz-Thouless transition temperature \cite{Kosterlitz:1973xp}.
We find that it becomes volume-independent after driving the system with the measurement processes $O^{(2)}$ and $O^{(3)}$ to its new equilibrium state, corresponding to a complete disordering of the spin system.
In order to study the transition between these two different phases, we again investigated the Binder ratio
\begin{equation}
 B_4(t)=\frac{\langle M^4\rangle(t)}{[\langle M^2\rangle(t)]^2} 
\end{equation}
and performed an accurate finite size analysis like for the anti-ferromagnetic Heisenberg model.
As shown in Fig.~\ref{fig:phase_transition_xy}, the various finite-volume curves for $B_4(t)$ do not intersect each other but their inflection points move to ever later times with increasing volumes.
Moreover, the magnetization density shows an exponential decay in time with an inverse decay rate $\gamma\tau=0.242(2)$.
All these observations indicate again that the phase transition does not occur at any finite point in time but is rather completed only after an infinite amount of time.

\begin{thebibliography}{100}

\bibitem{Stefanucci:2013}
  G.~Stefanucci and R.~van~Leeuwen,
  ``Nonequilibrium Many-Body Theory of Quantum Systems: A Modern Introduction'',
  Cambridge University Press (2013).

\bibitem{Berges:2004yj}
  J.~Berges,
  AIP Conf.\ Proc.\  {\bf 739} (2005) 3.

\bibitem{Aoki:2014}
 H.~Aoki, N.~Tsuji, M.~Eckstein, M.~Kollar, T.~Oka and P.~Werner,
 Rev.\ Mod.\ Phys.\ {\bf 86} (2014) 779.

\bibitem{White:1992zz}
  S.~R.~White,
  Phys.\ Rev.\ Lett.\  {\bf 69} (1992) 2863.

\bibitem{Schollwock:2005zz}
  U.~Schollw\"ock,
  Rev.\ Mod.\ Phys.\  {\bf 77} (2005) 259.

\bibitem{Cazalilla:2002}
  M.~A.~Cazalilla and J.~B.~Marston,
  Phys.\ Rev.\ Lett.\ {\bf 88} (2002) 256403.

\bibitem{Vidal:2004}
  G.~Vidal,
  Phys.\ Rev.\ Lett.\ {\bf 93} (2004) 040502.

\bibitem{White:2004}
  S.~R.~White and A.~E.~Feiguin,
  Phys.\ Rev.\ Lett.\  {\bf 93} (2004) 076401.

\bibitem{Verstraete:2004}
  F.~Verstraete, J.~J.~Garcia-Ripoll and J.~I.~Cirac,
  Phys.\ Rev.\ Lett.\ {\bf 93} (2004) 207204.

\bibitem{Daley:2004}
  A.~J.~Daley, C.~Kollath, U.~Schollw\"ock and G.~Vidal,
  J.\ Stat.\ Mech.\ (2004) P04005.

\bibitem{Buyens:2013yza}
  B.~Buyens, J.~Haegeman, K.~Van Acoleyen, H.~Verschelde and F.~Verstraete,
  Phys.\ Rev.\ Lett.\  {\bf 113} (2014) 091601.

\bibitem{Polkovnikov:2010}
 A.~Polkovnikov,
 Annals\ Phys.\ {\bf 325} (2010) 1790.

\bibitem{Romatschke:2005pm}
  P.~Romatschke and R.~Venugopalan,
  Phys.\ Rev.\ Lett.\  {\bf 96} (2006) 062302.

\bibitem{Gelis:2007kn}
  F.~Gelis, T.~Lappi and R.~Venugopalan,
  Int.\ J.\ Mod.\ Phys.\ E {\bf 16} (2007) 2595.

\bibitem{Gelis:2013rba}
  T.~Epelbaum and F.~Gelis,
  Phys.\ Rev.\ Lett.\  {\bf 111} (2013) 232301.

\bibitem{Berges:2013eia}
  J.~Berges, K.~Boguslavski, S.~Schlichting and R.~Venugopalan,
  Phys.\ Rev.\ D {\bf 89} (2014) 074011.

\bibitem{Kasper:2014uaa}
  V.~Kasper, F.~Hebenstreit and J.~Berges,
  Phys.\ Rev.\ D {\bf 90} (2014) 025016.

\bibitem{Parisi:1984cs}
  G.~Parisi,
  Phys.\ Lett.\ B {\bf 131} (1983) 393.

\bibitem{Huffel:1984mq}
  H.~Huffel and H.~Rumpf,
  Phys.\ Lett.\ B {\bf 148} (1984) 104.

\bibitem{Berges:2006xc}
  J.~Berges, S.~Borsanyi, D.~Sexty and I.-O.~Stamatescu,
  Phys.\ Rev.\ D {\bf 75} (2007) 045007.

\bibitem{Fukushima:2014iqa}
  K.~Fukushima and T.~Hayata,
  Phys.\ Lett.\ B {\bf 735} (2014) 371.

\bibitem{Feynman:1981tf}
  R.~P.~Feynman,
  Int.\ J.\ Theor.\ Phys.\  {\bf 21} (1982) 467.

\bibitem{Greiner:2002}
  M.~Greiner, O.~Mandel, T.~Esslinger, T.~W.~H\"ansch and I.~Bloch,
  Nature {\bf 415} (2002) 39.

\bibitem{Cirac:2008}
  J.~I.~Cirac and P.~Zoller,
  Nature\ Phys.\ {\bf 8} (2012) 264.

\bibitem{Bloch:2008}
  I.~Bloch, J.~Dalibard and S.~Nascimb\'{e}ne,
  Nature\ Phys.\ {\bf 8} (2012) 267.

\bibitem{Blatt:2008}
  R.~Blatt and C.~F.~Roos,
  Nature\ Phys.\ {\bf 8} (2012) 277.

\bibitem{AspuruGuzik:2008}
  A.~Aspuru-Guzik and P.~Walther,
  Nature\ Phys.\ {\bf 8} (2012) 285.

\bibitem{Houck:2008}
  A.~A.~Houck, H.~E.~T\"ureci and J.~Koch,
  Nature\ Phys.\ {\bf 8} (2012) 292.

\bibitem{Lewenstein:2012}
  M.~Lewenstein, A.~Sanpera and V.~Ahufinger,
  ``Ultracold Atoms in Optical Lattices: Simulating Quantum Many-Body Systems'',
  Oxford University Press (2012).

\bibitem{Kapit:2010qu}
  E.~Kapit and E.~J.~Mueller,
  Phys.\ Rev.\ A {\bf 83} (2011) 033625.

\bibitem{Zohar:2012ay}
  E.~Zohar, J.~I.~Cirac and B.~Reznik,
  Phys.\ Rev.\ Lett.\  {\bf 109} (2012) 125302.

\bibitem{Banerjee:2012pg}
  D.~Banerjee, M.~Dalmonte, M.~M\"uller, E.~Rico, P.~Stebler, U.-J.~Wiese and P.~Zoller,
  Phys.\ Rev.\ Lett.\  {\bf 109} (2012) 175302.

\bibitem{Banerjee:2012xg}
  D.~Banerjee, M.~B\"ogli, M.~Dalmonte, E.~Rico, P.~Stebler, U.-J.~Wiese and P.~Zoller,
  Phys.\ Rev.\ Lett.\  {\bf 110} (2013) 125303.

\bibitem{Zohar:2012xf}
  E.~Zohar, J.~I.~Cirac and B.~Reznik,
  Phys.\ Rev.\ Lett.\  {\bf 110} (2013) 125304.

\bibitem{Tagliacozzo:2012vg}
  L.~Tagliacozzo, A.~Celi, A.~Zamora and M.~Lewenstein,
  Annals Phys.\  {\bf 330} (2013) 160.

\bibitem{Tagliacozzo:2013}
  L.~Tagliacozzo, A.~Celi, P.~Orland, M.~W.~Mitchell and M.~Lewenstein,
  Nature\ Comm.\ {\bf 4} (2013) 2615.

\bibitem{Zohar:2013zla}
  E.~Zohar, J.~I.~Cirac and B.~Reznik,
  Phys.\ Rev.\ A {\bf 88} (2013) 023617.

\bibitem{Wiese:2013uua}
  U.-J.~Wiese,
  Annalen Phys.\  {\bf 525} (2013) 777.

\bibitem{Kuhn:2014rha}
  S.~K\"uhn, J.~I.~Cirac and M.~C.~Ba\~{n}uls,
  Phys.\ Rev.\ A {\bf 90} (2014) 042305.

\bibitem{Kossakowski:1972}
  A.~Kossakowski,
  Rep.\ Math.\ Phys.\ {\bf 3} (1972) 247.

\bibitem{Lindblad:1975ef}
  G.~Lindblad,
  Commun.\ Math.\ Phys.\  {\bf 48} (1976) 119.

\bibitem{Diehl:2008}
  S.~Diehl, A.~Micheli, A.~Kantian, B.~Kraus, H.~P.~B\"uchler and P.~Zoller,
  Nature\ Phys.\ {\bf 4} (2008) 878.

\bibitem{Prozen:2008}
  T.~Prosen and I.~Pi\v{z}orn,
  Phys.\ Rev.\ Lett.\ {\bf 101} (2008) 105701.

\bibitem{Verstraete:2009}
  F.~Verstraete, M.~M.~Wolf and J.~I.~Cirac,
  Nature\ Phys.\ {\bf 5} (2009) 633.

\bibitem{DallaTorre:2010}
  E.~G.~Dalla~Torre, E.~Demler, T.~Giamarchi and E.~Altman,
  Nature\ Phys.\ {\bf 6} (2010) 806.

\bibitem{Diehl:2010}
  S.~Diehl, A.~Tomadin, A.~Micheli, R.~Fazio and P.~Zoller,
  Phys.\ Rev.\ Lett.\ {\bf 105} (2010) 015702.

\bibitem{Muller:2012}
  M.~M\"uller, S.~Diehl, G.~Pupillo and P.~Zoller,
  Adv.\ Atom.\ Mol.\ Opt.\ Phys.\ {\bf 61} (2012) 1.

\bibitem{Lesanovsky:2013}
  I.~Lesanovsky, M.~van~Horssen, M.~Gu\c{t}\u{a} and J.~P.~Garrahan,
  Phys.\ Rev.\ Lett.\ {\bf 110} (2013) 150401.

\bibitem{Sieberer:2013}
  L.~M.~Sieberer, S.~D.~Huber, E.~Altman and S.~Diehl,
  Phys.\ Rev.\ Lett.\ {\bf 110} (2013) 195301.

\bibitem{Grandi:2013}
  C.~De~Grandi, A.~Polkovnikov and A.~W.~Sandvik,
  J.\ Phys:\ Cond.\ Matt.\ {\bf 25} (2013) 404216.

\bibitem{Banchi:2014}
  L.~Banchi, P.~Giorda and P.~Zanardi,
  Phys.\ Rev.\ E {\bf 89} (2014) 022102.

\bibitem{Lang:2014}
  N.~Lang and H.~P.~B\"uchler,
  arXiv:1408.4616 [quant-ph]

\bibitem{Evertz:1992rb}
  H.~G.~Evertz, G.~Lana and M.~Marcu,
  Phys.\ Rev.\ Lett.\  {\bf 70} (1993) 875.

\bibitem{Wiese:1994}
  U.-J.~Wiese and H.-P.~Ying,
  Z.\ Phys.\ B {\bf 93} (1994) 147.

\bibitem{Daley:2014}
  A.~J.~Daley
  Adv.\ Phys.\ {\bf 63} (2014) 77.

\bibitem{Banerjee:2014yea}
  D.~Banerjee, F.-J.~Jiang, M.~Kon and U.-J.~Wiese,
  Phys.\ Rev.\ B {\bf 90} (2014) 241104.

\bibitem{Schwinger:1960qe} 
  J.~S.~Schwinger,
  J.\ Math.\ Phys.\  {\bf 2} (1961) 407.

\bibitem{Keldysh:1964ud}
  L.~V.~Keldysh,
  Sov.\ Phys.\ JETP {\bf 20} (1965) 1018.

\bibitem{Konstantinov:1961}
  O.~V.~Konstantinov and V.~I.~Perel',
  Sov.\ Phys.\ JETP {\bf 12} (1961) 142.

\bibitem{Griffiths:1984rx}
  R.~B.~Griffiths,
  J.\ Stat.\ Phys.\  {\bf 36} (1984) 219.

\bibitem{Beard:1996wj}
  B.~B.~Beard and U.-J.~Wiese,
  Phys.\ Rev.\ Lett.\  {\bf 77} (1996) 5130.

\bibitem{Gockeler:1990ik}
  M.~G\"ockeler and H.~Leutwyler,
  Phys.\ Lett.\ B {\bf 253} (1991) 193.

\bibitem{Gockeler:1990zn}
  M.~G\"ockeler and H.~Leutwyler,
  Nucl.\ Phys.\ B {\bf 350} (1991) 228.

\bibitem{Hasenfratz:1993}
  P.~Hasenfratz and F.~Niedermayer,
  Z.\ Phys.\ B {\bf 92} (1993) 91.

\bibitem{Sandvik:2010}
  A.~W.~Sandvik and H.~G.~Evertz,
  Phys.\ Rev.\ B {\bf 82} (2010) 024407.

\bibitem{Gerber:2009rd}
  U.~Gerber, C.~P.~Hofmann, F.-J.~Jiang, M.~Nyfeler and U.-J.~Wiese,
  J.\ Stat.\ Mech.\  {\bf 2009} (2009) P03021.

\bibitem{Hohenberg:1977ym}
  P.~C.~Hohenberg and B.~I.~Halperin,
  Rev.\ Mod.\ Phys.\  {\bf 49} (1977) 435.

\bibitem{Binder:1981zz}
  K.~Binder,
  Phys.\ Rev.\ Lett.\  {\bf 47} (1981) 693.

\bibitem{Binder:1984}
  K.~Binder and D.~P.~Landau,
 Phys.\ Rev.\ B\  {\bf 30} (1984) 1477.

\bibitem{Bietenholz:1995zk}
  W.~Bietenholz, A.~Pochinsky and U.-J.~Wiese,
  Phys.\ Rev.\ Lett.\  {\bf 75} (1995) 4524.

\bibitem{Chandrasekharan:1999cm}
  S.~Chandrasekharan and U.-J.~Wiese,
  Phys.\ Rev.\ Lett.\  {\bf 83} (1999) 3116.

\bibitem{Kawashima:1995a}
  N.~Kawashima and J.~E.~Gubernatis,
  J.\ Stat.\ Phys.\ {\bf 80} (1995) 169.

\bibitem{Kawashima:1995b}
  N.~Kawashima and J.~E.~Gubernatis,
  Phys.\ Rev.\ E {\bf 51} (1995) 1547.

\bibitem{Evertz:2000rk}
  H.~G.~Evertz,
  Adv.\ Phys.\  {\bf 52} (2003) 1.

\bibitem{Jiang:2011}
  F.-J.~Jiang,
  Phys.\ Rev.\ B\ {\bf 83} (2011) 024419.

\bibitem{Gerber:2011ya}
  U.~Gerber, C.~P.~Hofmann, F.-J.~Jiang, G.~Palma, P.~Stebler and U.-J.~Wiese,
  J.\ Stat.\ Mech.\  {\bf 1106} (2011) P06002.

\bibitem{Kosterlitz:1973xp}
  J.~M.~Kosterlitz and D.~J.~Thouless,
  J.\ Phys.\ C {\bf 6} (1973) 1181.

\end{thebibliography}
\end{document}